\documentclass[aps,prb,twocolumn,superscriptaddress]{revtex4-2}

\usepackage{graphicx} 
\usepackage{graphics}
\usepackage{epsfig}
\usepackage{appendix}
\usepackage{amsmath, amssymb, braket}
\usepackage{dirtytalk}
\usepackage{hyperref, listings}
\usepackage{bbold}
\usepackage{esvect, cancel}
\usepackage{svg}
\usepackage{comment}
\usepackage{float}
\usepackage{mathtools}
\usepackage{enumitem} 

\begin{document}

\title{Dynamics of Many-Emitter Ensembles: Probing Cooperative Evolution with Scalable Quantum Circuits}

\author{Vincent Iglesias-Cardinale}
\email{vpiglesi@buffalo.edu}
\affiliation{Department of Physics, University at Buffalo SUNY, Buffalo, New York 14260, USA}
\author{Shreekanth S. Yuvarajan} 
\affiliation{Department of Physics, University at Buffalo SUNY, Buffalo, New York 14260, USA}
\author{Herbert F. Fotso}
\email{hffotso@buffalo.edu}
\affiliation{Department of Physics, University at Buffalo SUNY, Buffalo, New York 14260, USA}

\begin{abstract}
\noindent Many-particle quantum systems often give rise to exotic behaviors in their nonequilibrium dynamics that are rather challenging to reveal with analytical methods or with classical computation.  
Here, we consider the case of a system of many quantum emitters coupled through a radiation bath. By adopting an efficient mapping of the bosonic modes onto a set of quantum bits, we implement quantum circuits, compatible with NISQ (Noisy Intermediate-Scale Quantum) era systems, that allow us to investigate the dynamics of the ensemble as a function of various parameters, including the number of emitters, the spectral inhomogeneity in the system, the emission lifetime of independent emitters, and the spatial separation between emitters. The quantum algorithms afford us the capacity to precisely track the emergence of cooperative dynamics, manifested through superradiant emission, as the system is tuned towards optimal coupling with respect to various parameters. We are particularly able to characterize superradiant emission in an inhomogeneous ensemble as a function of the linewidth of the individual emitters.
These quantum algorithms avoid approximations performed in conventional studies of many-emitter systems and provide a robust and intuitive characterization. Despite being limited to a small number of qubits, the present calculations are found to provide a reliable characterization validated by comparison with analytical solutions and classical computation results in their respective regimes of validity. 
These findings indicate that the approach can be employed to effectively simulate a broad variety of many-emitter systems.
\end{abstract}

\maketitle

\section{Introduction}
\label{sec:Introduction}

\noindent The simulation of many-particle systems remains an outstanding problem in physics in general. Particularly, the challenge of obtaining the time-dependent quantum state of a many-particle physical system arises in various sub-fields of physics, with a broad range of potential applications. Due to the intrinsic complexity of large quantum systems, the analytical treatment often requires approximations that subsequently limit the applicability of the obtained solutions. To this effect, a variety of computational methods have been developed over time. However, due to the exponential scaling of the Hilbert space for many-particle quantum systems, clear limits exist in the usage of classical numerical methods. These limits are not expected to be addressable by the historically documented growth of computational power of the classical type. Indeed, a great deal of the recently achieved progress in the field is due to the continued investments in the formulation of novel, efficient and more powerful algorithms~\cite{ALPS2, TRIQS, QE-2017, COMSOL, Ansys, qutip2025}. Besides these challenges, the deployment of classical computing algorithms is often paired with an additional layer of complexity arising between the physical system and its model, and from sophisticated numerical implementations. Such abstractions are more error prone and may render the computation rather difficult to track.

While the dynamics of ensembles of quantum emitters and the emergence/destruction of cooperative behaviors have been investigated with great interest for several decades now~\cite{Dicke1954, GrossHaroche1982, BonifacioPRA1971_I, BonifacioPRA1971_II}, interest in the question remains strong in part as a result of new applications in various areas ranging from cavity quantum electrodynamics~\cite{RaimondEtAlPRL1982, SlamaEtAlPRL2007,Delanty_2011, BaskicPRL2014} to quantum sensing~\cite{Colombo2022, Franke2023, Brinke2015, DeMille2024, KoppenhoeferPRL2023}. Despite this sustained attention, theoretical characterization is still limited to approximate analytical methods, often using mean-field methods, or to classical simulations of few particle systems~\cite{BonifacioPRA1971_I, BonifacioPRA1971_II, MasonEtAlPRL2020, ClemensEtAl2003, MotAsenjo-GarciaEtAlPRL2023, HosseinabadiOksanaMarinoPRXquantum2025}.

Progress in the construction of quantum computing platforms, although still limited to NISQ (Noisy Intermediate-Scale Quantum) systems, has brought about the promise of a novel paradigm for the simulation of many-particle systems~\cite{Preskill2018quantumcomputingin, Fauseweh2024, DaleyZoller_Nature2022, BhartiAspuru-Guzik_RMP2022, IppolitiKhemani_prx2021, FotsoSPIE2025, yuvarajan2025}. In the present paper, we adopt an effective mapping of bosonic modes of radiation onto a set of quantum bits and we implement quantum circuits for the simulation of an ensemble of many emitters coupled through a bath of radiation modes. 
Due to the robustness of these quantum algorithms~\cite{FotsoSPIE2025, yuvarajan2025}, we are able to track the emergence of cooperative dynamics manifested by superradiant emission from the ensemble as a function of various system parameters including the spectral inhomogeneity in the system, the number of emitters, the emission lifetime of independent emitters, and the spatial separation between emitters in the ensemble.

Here, we avoid integrating out the bosonic bath and explicitly track the dynamics of the entire system: emitters and bosonic modes of the radiation. In this way, our quantum circuits, by mapping the model onto a system of qubits, provide a clear and intuitive characterization of the many-emitter system. This paper presents an approach that can be reliably applied to a broader range of light-matter interaction problems and particularly to those of interest in quantum information processing~\cite{FotsoEtal_PRL2016, Fotso_noisyTLS2022, Fotso_TPI_PRB_2019, Lei2023}. 

We note that for bosonic systems, quantum computation might seem out of reach on present hardware due to the fact that, unlike their fermionic counterparts, the number of excitations in a given mode, or mode occupation, is in principle unlimited. However, for the problem at hand, we introduce a mapping of bosonic modes onto qubits. We find that with efficient qubit mapping of the multi-photon excitation of the same mode, all characteristic time and energy scales of a superradiant homogeneous or inhomogeneous ensemble of several atoms can be well captured with a maximum of $\sim$20 qubits. This is in part because the system can reasonably be considered isolated during the timescale of the simulation (typically on the order of the relaxation time of independent emitters) and the preparation of the initial state caps the number of  excitations in the system to a manageable threshold. 

The rest of the paper is structured as follows. In section \ref{sec:Model}, we present the model for the system of many emitters in a radiation bath. In section~\ref{sec:Methods}, we discuss the construction of the qubit Hamiltonian as well as the quantum circuit to simulate the dynamics after discretizing the time axis via a Suzuki-Trotter decomposition. We also provide a characterization of the systematic errors in this solution. In section ~\ref{sec:Results}, we present our results for the dynamics of the system as a function of time. This includes tracking the occupation of the radiation modes, the overall radiation intensity, as well as its dependence on the number of emitters. In this section, we emphasize the case of an inhomogeneous ensemble, confirming that our algorithm shows insight that cannot be obtained from crude approximations or mean-field treatments. We end with our conclusions in section~\ref{sec:Conclusion}.
A number of details are provided in the appendices. Namely, a validation of our algorithm by reproducing well established results and a discussion of the emitted radiation from spatially diffuse ensembles is presented in Appendix~\ref{app:benchmarking}, an appraisal of the compatibility of the algorithm with existing NISQ systems can be found in Appendix~\ref{app:hardware}, and a proof of the recursive relation developed for bosonic operators in the binary representation is presented in Appendix~\ref{app: recursion relation}.\\

\section{Model}
\label{sec:Model}

\noindent We consider a system of $N_A$ two-level atoms coupled to a common radiation bath and described by the many atom extension of the Weisskopf-Wigner Hamiltonian~\cite{WeisskopfWigner1930, Dicke1954, GrossHaroche1982}:
\begin{eqnarray}
        H &=& \frac{1}{2}\sum_\alpha\omega_\alpha\sigma_\alpha^z +\sum_{\mathbf k\lambda}\omega_{\mathbf k} a_{\mathbf k\lambda}^\dagger a_{\mathbf k\lambda} \nonumber \\
        &-&i\sum_\alpha \sum_{\mathbf k\lambda}g_{\mathbf k\lambda}^\alpha(\sigma_\alpha^+ + \sigma_\alpha^-)\left(e^{-i\mathbf k\cdot\mathbf r_\alpha}a_{\mathbf k\lambda}^\dagger - e^{i\mathbf k\cdot\mathbf r_\alpha}a_{\mathbf k\lambda}\right). \nonumber \\
        & &
\label{eq:Hamiltonian1}
\end{eqnarray}
Here, the operators $\sigma^z_{\alpha}=|e_{\alpha}\rangle\langle e_{\alpha}|-|g_{\alpha}\rangle\langle g_{\alpha}|$, $\sigma^+_{\alpha} =|e_{\alpha}\rangle\langle g_{\alpha}|$, and $\sigma^-_{\alpha} = |g_{\alpha}\rangle\langle e_{\alpha}| = (\sigma^+_{\alpha})^\dagger$ are, respectively, the $z$-axis Pauli matrix,
the raising, and the lowering operators for the two-level atom $\alpha$, with $|g_{\alpha}\rangle$ its ground and $ |e_{\alpha}\rangle$ its excited state. 
$a_{k\lambda}$ ($a^{\dagger}_{k\lambda}$) is the annihilation (creation) operator of the $k$-th photon mode with polarization $\lambda$, $g^{\alpha}_{k \lambda}$ is its coupling strength to atom $\alpha$, the position of which is $\mathbf{r}_{\alpha}$.
The free emission rate of atom $\alpha$ is $\Gamma_{\alpha}$, related to it's free emission time $\tau_\alpha$ when emitting independently by $\tau_\alpha = 1/\Gamma_\alpha$.  Throughout this paper we consider identical free emission rates for all atoms in a given ensemble regardless of their natural emission frequency, allowing for the simplification $\Gamma_\alpha = \Gamma_0$ and $g_{\mathbf k\lambda}^\alpha = g_{\mathbf k\lambda} = g$ for all atom-mode couplings. Consequently, all times are presented in units of $1/\Gamma_0$. When the number of bosonic modes $N$ is reduced to one, for a single atom, we obtain the Jaynes-Cummings~\cite{JaynesCummings} model describing a two-level system coupled to a cavity.

The general quantum state of the \textit{multi-atom + photons} system can be written as: 
\begin{equation}
|\psi(t)\rangle = \sum_i C_i |\psi^i _{\text{atoms}}(t)\rangle \otimes |\psi^i_{\text{field}}(t)\rangle ,    
\end{equation}
where $C_i$ is a complex number, $|\psi^i _{\text{atoms}}(t)\rangle$ is the quantum state of the atoms, while $|\psi^i_{\text{field}}\rangle$ is the state of the field modes. For $N_A$ atoms $|\psi^i _{\text{atoms}}(t)\rangle$ lives in a $2^{N_A}$ dimensional space.  
The size of the Hilbert space for the field is similarly $2^{\sum_k q_k}$, where $q_k$ represents the number of qubits allocated to modeling mode $k$, and the sum is over all field modes. We implement a binary mapping similar to that of Ref.~\onlinecite{huang2022qubitizationbosons} so that $2^{q_k} -1$ excitations may be modeled with only $q_k$ qubits, and therefore the field Hilbert space grows as $2^{\sum_k \lceil\log_2(N_k+1)\rceil}$, where $N_k$ is the number of excitations allowed in mode $k$, and $\lceil\cdot\rceil$ represents the ceiling function.

We are interested in the dynamics under the Hamiltonian (\ref{eq:Hamiltonian1}) of the system when it is initialized with all atoms in their excited states and all radiation modes empty: $|\psi(0)\rangle = |1\rangle^{\otimes N_A}\otimes |0\rangle^{\otimes \sum_k q_k}$.

Since our algorithm allows us to evolve the state of the entire system, we extract the occupation of individual bosonic modes  as a function of time by taking the expectation value of the number operator for mode $k$, $\langle n_k(t)\rangle = \langle\psi(t)|a_k^\dagger a_k|\psi(t)\rangle$. The total mode occupation is simply the sum of these contributions over all modes $\langle n(t)\rangle = \sum_k \langle n_k(t)\rangle$.

\noindent The intensity of the radiation produced through the relaxation of the atoms, transferring the excitation from the $N_A$ atoms into the radiation modes, is the number of photons per unit time emitted by the ensemble and it is given by: 
\begin{eqnarray}
    I_{N_A}(t) &= &\Gamma_0 \langle S^+S^-\rangle \label{eq:intensity1} \\
    & = &  \Gamma_0\langle\sum_{\alpha\beta}\sigma_\alpha^+\sigma_\beta^-\rangle \label{eq:intensity2} \\ 
    & = & \Gamma_0\sum_{\alpha\neq \beta}\langle\sigma_\alpha^+\sigma_\beta^-\rangle + \Gamma_0\sum_\alpha\langle\sigma_\alpha^+\sigma_\alpha^-\rangle
    \label{eq:intensity3} \\
        & = & I_{N_A\text{,C}}(t) + I_{N_A\text{,NC}}(t) \label{eq:coherentNonCoherent},
\end{eqnarray}

\noindent In Eq.(\ref{eq:intensity1}), $S^+ = \sum_{\alpha} \sigma_{\alpha}^+$ and $S^- = \sum_{\alpha} \sigma_{\alpha}^-$ and in Eq.(\ref{eq:coherentNonCoherent}), the intensity is separated into its cooperative and its non-coherent contributions, defined respectively by
\begin{equation}
    I_{N_A\text{,C}}(t) = \Gamma_0\sum_{\alpha \neq  \beta}\langle \sigma_{\alpha}^+\sigma_{\beta}^-\rangle
    \label{eq:coherence}
\end{equation}
and 
\begin{equation}
    I_{N_A\text{,NC}}(t) = \Gamma_0\sum_{\alpha}\langle\sigma_{\alpha}^+\sigma_{\alpha}^-\rangle.
\end{equation}

In the absence of cooperative dynamics, each atom emits independently, due only to its own coupling to the radiation field. In this case the coherence, $I_{N_A,\text{C}}(t)$, vanishes for the duration of the emission process. The ensemble intensity profile in this case is simply the incoherent sum of the exponential intensity profiles from all $N_A$ atoms.
Cooperative dynamics can induce superradiance which is marked by a build up of coherence, $I_{N_A\text{,C}}(t)$, during the emission process, leading to a burst in the intensity profile from the ensemble.

\section{Methods}
\label{sec:Methods}

\noindent In general, the solution of the problem when tracking the bosonic modes and the atomic states is constrained by the size of the Hilbert space and the solution through classical simulations can only be performed for small systems. Alternatively, a master equation can be derived for the evolution of the density matrix operator of the atoms and either solved in a mean-field for special cases or integrated numerically as can be done with the QuTiP libraries~\cite{qutip2025}.
In this paper, we present an alternative avenue that aims to eventually leverage the growing quantum computing infrastructure for a more scalable solution. Note that although the results presented in the current paper are obtained on a quantum simulator, they offer a robust formulation to track the dynamics of the entire system (atoms and bosonic modes). A discussion of the product of these algorithms on publicly available quantum hardware at the time of publication can be found in Appendix~\ref{app:hardware}. \\
 
The quantum simulation of the many-atom extension of the Weisskopf-Wigner Hamiltonian is first dependent on the analytic procedure to ``qubitize" the Hamiltonian i.e. to map the operators onto quantum gates. After the qubitization, time-evolution is accomplished by discretizing the time axis using a Suzuki-Trotter decomposition. A quantum circuit is then built as a sequence of single and two-qubit gates on the appropriate qubits, representing time-evolution of the system from $t=0$ (where the qubits are initialized in the appropriate states) to some final time $t=t_f$ (where the qubit states are measured). This process is repeated until a sufficient amount of data is collected at $t_f$. To capture the time dynamics, this entire process is repeated iteratively using an incrementally increasing $t_f$ until the desired maximum evolution time, $T$, is achieved. 

In this work, except in Appendix~\ref{app:hardware}, we used a quantum simulator in Qiskit, so the repeated trials are not necessary~\cite{Qiskit2024}. Once the states are extracted at a certain time, the expectation values of any observable may be found at that time. In this way we track photon occupation in the modes, the number of excited atoms, the emission intensity, etc.

\subsection{Qubit Hamiltonian}
\label{subsec:qubitHamiltonian}
\noindent The first step in designing the quantum algorithm is to construct the ``qubits-only" Hamiltonian from the original ``\textit{atoms+field}" Hamiltonian (\ref{eq:Hamiltonian1}). If we assume a maximum of one excitation in each bosonic mode, we can use a reverse Holstein–Primakoff type transformation~\cite{HP_Transform}  with the relations:
\begin{eqnarray}
\sigma_{\alpha}^z &=& -Z_{\alpha} \nonumber \\
\sigma_{\alpha}^+ &=& \frac{1}{2}\left(X_{\alpha} -iY_{\alpha}\right) \nonumber \\
\sigma_{\alpha}^- &=& \frac{1}{2}\left(X_{\alpha} + iY_{\alpha}\right) \nonumber \\
a^\dagger_k &\equiv& \sigma^+_k = \frac{1}{2}(X_k - iY_k) \nonumber \\
a_k &\equiv& \sigma^-_k = \frac{1}{2}(X_k + iY_k).
\label{eq:qubitization1}
\end{eqnarray}
Here, $X, \; Y, \; Z,  \mathrm{and} \; I $ are Pauli operators on the qubits and we use the subscripts $\alpha$ and $k$ to identify qubits corresponding to the atoms and the radiation modes respectively. 
Substituting these into the Hamiltonian (\ref{eq:Hamiltonian1}), we would get the ``qubitized" Hamiltonian:
\begin{eqnarray}
        H &=& -\frac{1}{2}\sum_\alpha\omega_\alpha Z_\alpha +\frac{1}{2}\sum_{k}\omega_{k}(I_k-Z_k) \nonumber \\
        &-&\sum_\alpha \sum_{k}g_k\bigg\{\cos(\mathbf k \cdot \mathbf r_\alpha)X_\alpha Y_k  \nonumber \\
        &+&\sin(\mathbf k \cdot \mathbf r_\alpha)X_\alpha X_k\bigg\}
\label{eq:HamiltonianQubitized}
\end{eqnarray}

\noindent As mentioned above, the mappings of Eqs.(\ref{eq:qubitization1}) account for only a single excitation of each bosonic mode. As we consider ensembles of emitters coupled to a shared bath of bosonic modes, we must account for occupations greater than unity in the modes most likely to be occupied. To this end, to those modes for which a greater occupation is expected, we allocate multiple qubits in a binary representation~\cite{veis2016quantumchemistrybornoppenheimerapproximation, Kottmann_2021, huang2022qubitizationbosons, peng2023quantumsimulationbosonrelatedhamiltonians, escofet2024revisitingmappingquantumcircuits, jiang2025binary} that allows for $2^{q_k}-1$ excitations in mode $k$, where $q_k$ is the number of qubits allocated to that mode. The mapping of the bosonic operators must be updated to match this representation. We therefore develop the following recursion relation from which the creation operator in the binary representation can be obtained:
\begin{eqnarray}
        a_{q_k}^\dagger(c) \nonumber
        & = & \frac{I+Z}{2}\otimes a_{q_k-1}^\dagger(c) \\
        & + & \frac{\sqrt{c + 2^{q_k - 1}}}{2^{q_k}}(X-iY)\otimes(X+iY)^{\otimes(q_k - 1)}\nonumber \\
        & + & \frac{I-Z}{2}\otimes a_{q_k-1}^\dagger(c + 2^{q_k - 1}).
    \label{eq:summaryRecursionRelation}
\end{eqnarray}

\noindent The creation operator given by Eq.(\ref{eq:summaryRecursionRelation}) acts on mode $k$ in the usual way. The parameter $c$ represents an addition to the number of photons counted by the number operator in this representation, it is necessary only for the construction of the operator and should be set to zero for general use. With the single qubit allocation ($q_k=1$),  $a_1^\dagger(0)\equiv\frac{1}{2}(X_k-iY_k)$, as in Eqs.(\ref{eq:qubitization1}). For a detailed discussion and derivation of $a_k^\dagger$, $a_k$ and $n_k$ in this represention, we refer the reader to Appendix~\ref{app: recursion relation}.

\subsection{Time Evolution, Quantum Circuits}
\label{subsec:circuit}

\noindent The system is initialized in a state with all atoms in their excited states and all radiation modes empty. Since our goal is to track the dynamics of the system, we want to evolve it in time starting from these initial conditions. To this end, we discretize the time axis into finite time steps $\delta t$ and apply the Suzuki-Trotter decomposition on the time-evolution operator for the system. Under the Hamiltonian (\ref{eq:HamiltonianQubitized}), the time evolution is written as:
\begin{equation}
  U(0, t) \approx  \left( \mathrm{e}^{-\mathrm{i} H_0 \delta t} \mathrm{e}^{-\mathrm{i}H_{int} \delta t} \right)^{t/\delta t}
\end{equation}
and
\begin{equation}
U(t, t+\delta t) \approx \mathrm{e}^{-\mathrm{i}H_0 \delta t} \mathrm{e}^{-\mathrm{i}H_{int} \delta t}  
\label{eq:trotterization}
\end{equation}
with
\begin{eqnarray}
H_0 &=& -\frac{1}{2}\sum_\alpha\omega_\alpha Z_\alpha +\frac{1}{2}\sum_{k}\omega_{k}(I_k-Z_k)  \\
H_{int} &=&-\sum_\alpha \sum_{k}g_k\bigg\{\cos(\mathbf k \cdot \mathbf r_\alpha)X_\alpha Y_k  \nonumber \\
        &+&\sin(\mathbf k \cdot \mathbf r_\alpha)X_\alpha X_k\bigg\}
\end{eqnarray}

\begin{figure}[t] 
  \centering
      	\includegraphics[width=8.0cm]{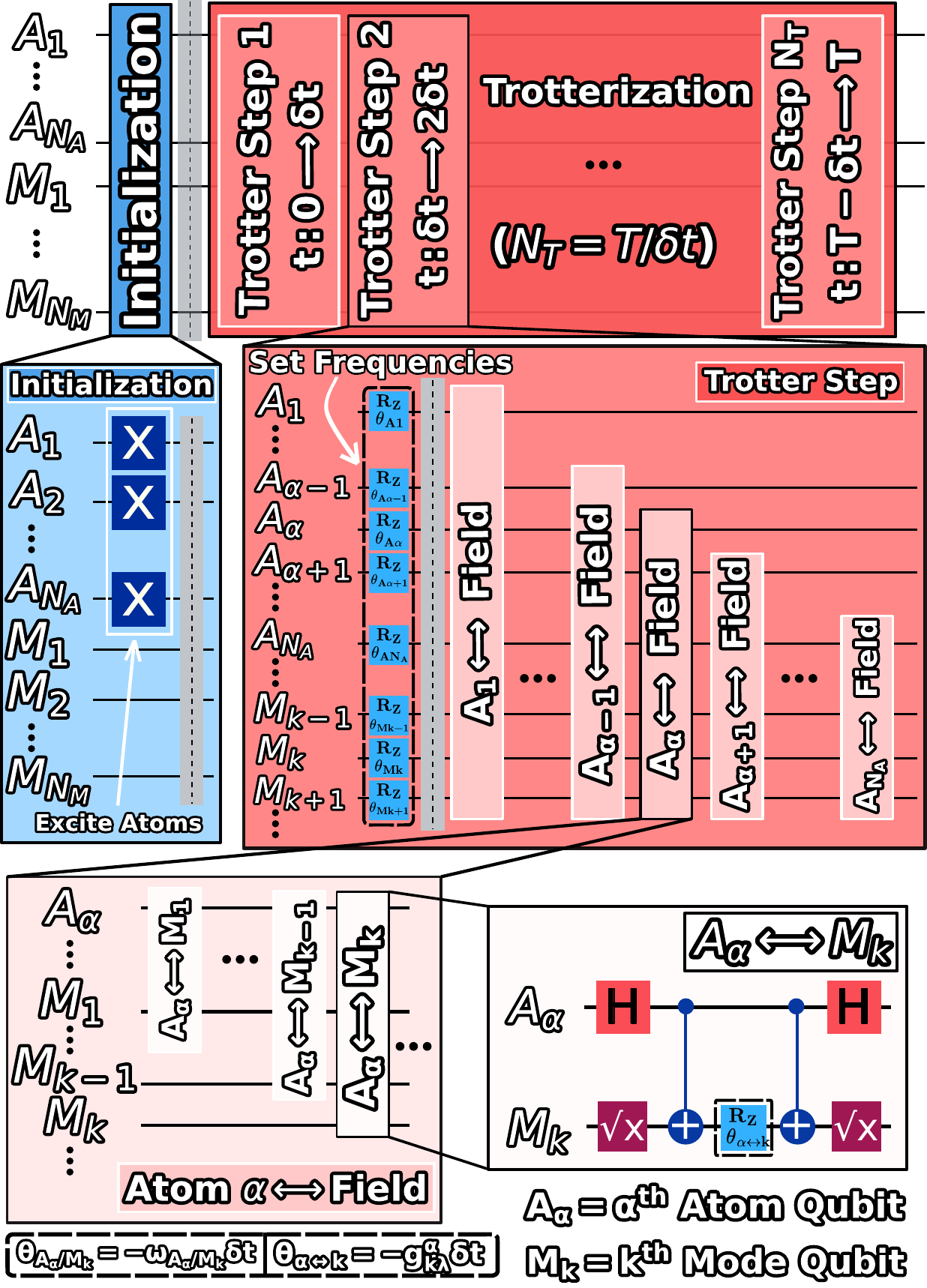}
      	\caption{Illustration of the circuit to simulate the time-evolution of the many atom + radiation system under Hamiltonian (\ref{eq:HamiltonianQubitized}).
        }
        \label{fig:circuit}
\end{figure}

\noindent The number of time steps is chosen to minimize the systematic error while allowing simulations up to multiple lifetimes of the individual emitters.
The initialization and the time evolution are illustrated in Fig.~\ref{fig:circuit}. The figure shows a circuit with 
$N_A + N_M$ qubits with $N_A$ qubits representing the atoms and $N_M = \sum_k q_k$ qubits representing the bosonic modes. The circuit first initializes the qubits in the states corresponding to all atoms in the excited state and all radiation modes empty.

During each Trotter step of the time evolution, appropriate phases are applied to each of the qubits, corresponding to the frequency of the atom or of the radiation mode, and the circuit performs the appropriate sets of two qubit gates that effectively couple the atoms and the radiation modes. The results presented in this paper are obtained by implementing this quantum algorithm in a Python code using the Qiskit library~\cite{Qiskit2024}.

\subsection{Systematic Error Analysis}
\label{subsec:SystematicErrors}
\noindent The quantum circuits developed above allow us to reliably evolve the many-emitter system from its initial configuration up to an arbitrary time. It is however important to assess the systematic errors accrued during the time evolution with this algorithm. The main sources of systematic errors are, on the one hand, the limited number of bosonic modes and the limited number of atoms
that can be appropriately treated on existing classical or quantum computing platforms, and, on the other hand, the error due to the finite Trotter time step. We also highlight some of the challenges arising from the gate fidelity on the quantum hardware in Appendix~\ref{app:hardware}. 

\subsubsection{Number of modes and atoms, limited mode occupation}
\noindent While we are able to observe superradiance in up to seven atoms with our encoding scheme using just 20 qubits, there are several limitations that arise from the constrained number of available qubits. To illustrate the limitation of the number of radiation modes, consider the Weisskopf-Wigner model describing a single two-level atom and many radiation modes. This model appropriately describes the spontaneous emission of the atom into the radiation bath. If we simplify this model to only include a single radiation mode, we are left with the Jaynes-Cummings model corresponding to a two-level atom coupled to a single-mode cavity, the dynamics of which is characterized by Rabi oscillations where the excitation periodically oscillates between the atom and the radiation mode. This is well captured by the algorithm presented in the current paper (see Figure~\ref{fig:intensityInhomogeneousEnsemble}). If we simulate the system with a gradually increased number of modes, 
we observe that the re-excitation of the atom is delayed to a time that is proportional to the number of modes. In fact, spontaneous emission corresponds to the case of an infinite number of modes and is thus difficult to capture within the finite limitations of this, or any, discretized model. 

In our simulations, the coupling strengths $g_k$ to the radiation modes are chosen to appropriately capture the relaxation time of the individual atoms. Fermi's Golden Rule relates the relaxation rate $\Gamma_0$, to the coupling coefficients for a two-level atom coupled to a reservoir with evenly spaced discrete energy levels separated by energy $\delta$~\cite{Cohen_Tannoudji_Book1992},
\begin{equation}
    \Gamma_0 = \frac{2\pi}{\hbar}g_{\mathbf k\lambda}^2\delta. 
\end{equation} 
This, as noted above, allows for the re-excitation at long times of the atoms. Nevertheless, we find that  with a sufficient though finite number of modes we are able to reliably characterize the decay of ensembles of several atoms.  

A further constraint on the system size is related to the number of atoms and the maximum number of excitations in a given mode. While the former is a system parameter that will be tuned to explore the emergence of cooperative dynamics, the latter has to be addressed in an adequate manner. A priori, one may want to allow for the maximum occupation of each mode to be equal to the total number of atoms in the system. But, due to the statistical spread of the emitted radiation, this is, in general, not necessary. Each radiation mode requires at least one qubit. In general, the photon capacity of a radiation mode modeled with $q_k$ qubits in the binary representation is $2^{q_k}-1$. However, we can often use the emission lineshape to optimize the number of qubits allocated to each mode. For instance, for a system with only a handful of atoms in the ensemble, as long as there are multiple modes within $\omega_\alpha\pm\Gamma_0$, since even the resonant mode occupation is unlikely to surpass 3, 
the modes can each be appropriately described with just one or two qubits. As expected, this will have to be adjusted as the number of atoms is increased.
One additional aspect of the model is that our radiation modes will be centered around the emission frequency of our emitters and its frequency range will be limited, since we are capped in the number of modes that we may include in the simulation. This limited spectral width will result in inaccuracies in the short-time dynamics in the simulation that can be mitigated by systematically tuning the spectral width.     

The limitations identified in this section will potentially be addressed by scalable quantum systems of the future. It is however remarkable to highlight how well the present implementations are able to capture irreversible decay even when limited to $\sim20$ qubits. 

\subsubsection{Suzuki-Trotter time step error}

\noindent The choice of the magnitude of the time step in the discretization of the time axis has to be made judiciously. On the one hand, we want to minimize the error incurred in the Suzuki-Trotter decomposition. On the other hand, we want to be able to track the dynamics of the system up to long enough times to be able to capture the appropriate timescales of the system. Although the results in the present manuscript are obtained on a quantum simulator with perfect two-qubit gate fidelities, we note that the number of Trotter steps is also proportional to the number of two-qubit gates performed in the time evolution up to a target maximum time further emphasizing why a small number of Trotter steps is desired. In our implementation, the Trotter steps are chosen by monitoring the accuracy of the calculation through the conservation of the total energy since the system is isolated. Once the Trotter step is sufficiently small, decreasing it further will have little to no effect on the dynamics of the system over the evolution time.

\begin{figure}[t] 
  \centering
        \includegraphics[width=\columnwidth]{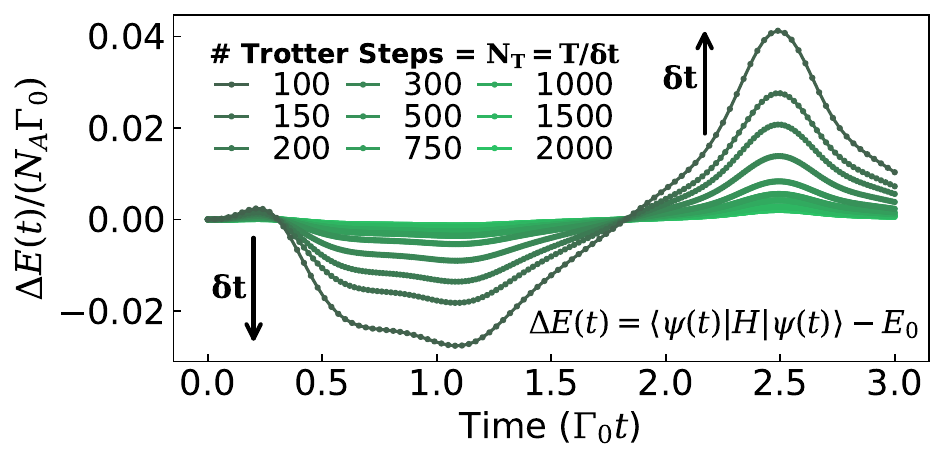}
      	\caption{Assessing the conservation of energy in the system as a function of time for various values of Trotter step size, $\delta t$. The system consists of four resonant atoms in the long wavelength limit in a bath of seven modes.
        }
	\label{fig:trotterError}
\end{figure}

By tracking energy conservation as a function of time, Figure~\ref{fig:trotterError} illustrates the systematic errors in the simulation of a system of 4 resonant atoms, in the long wavelength limit, in a bath of seven radiation modes as a function of the Trotter time step, 
$\delta t = T/N_T$, where $T$ and $N_T$ are the total evolution time and the number of Trotter steps, respectively. We find that, even for only 100 Trotter steps over three lifetimes, $3\tau$ with  $\tau=1/\Gamma_0$, energy is conserved to well within five percent of $N_A\Gamma_0$, a characteristic energy scale for this system.

\section{Results}
\label{sec:Results}

\noindent We apply the binary mapping to the \textit{multi-atom + photons} Hamiltonian (\ref{eq:Hamiltonian1}) and track the dynamics of the system in a variety of settings to assess different properties, and examine the emergence of cooperative relaxation from the ensemble. Note that unlike master equation solutions where the radiation modes are integrated out and only the atomic degrees of freedom are tracked through their density matrix operator, the quantum circuits enable detailed characterization of the entire system including the radiation modes. In the analysis of our results, when appropriate, we will make comparisons with those of the numerical solution of the master equation using the QuTiP libraries~\cite{qutip2025}.
We first examine the dynamics of a spectrally homogeneous ensemble to benchmark our algorithm before moving to the spectrally inhomogeneous ensemble.

\begin{figure}[t] 
\includegraphics[width=\columnwidth]{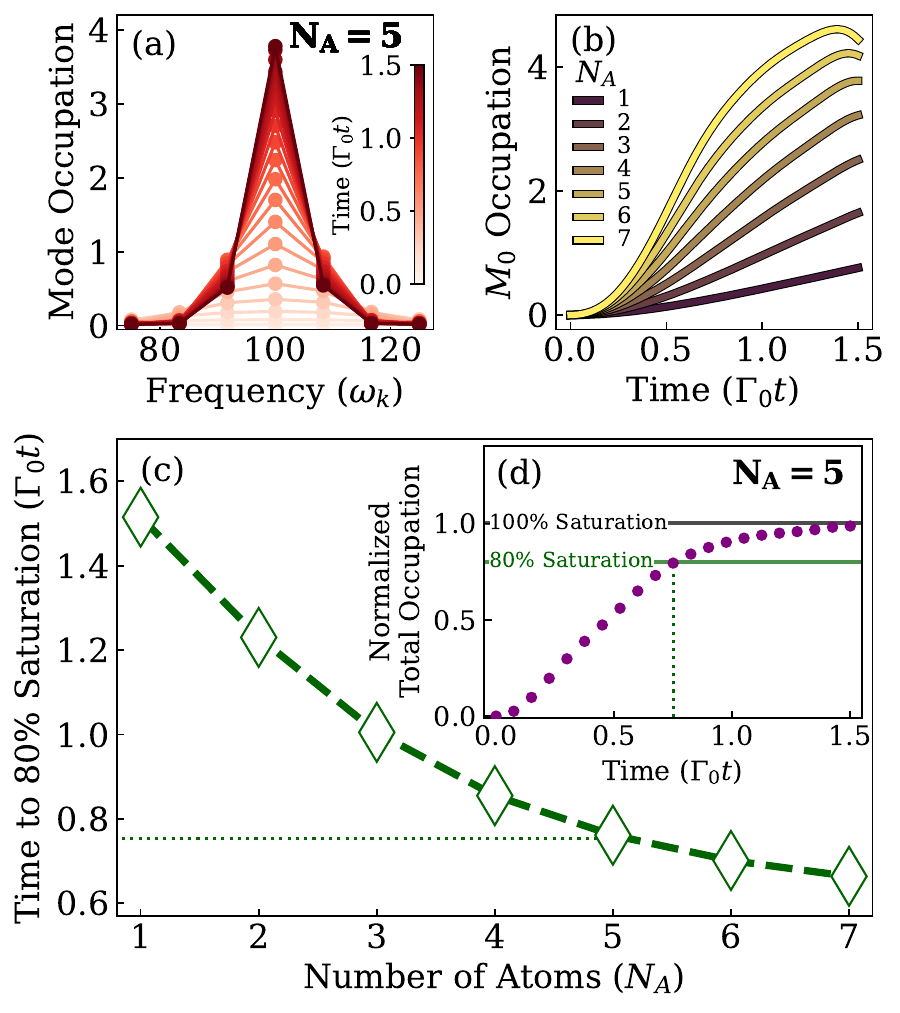}
\caption{\textbf{(a)} Emission spectrum of a system of 5 resonant atoms in a bath of 7 radiation modes in the long wavelength limit, shown as a function of time with darker shades representing later times. 
\textbf{(b)} Occupation of the resonant mode ($\omega_{M_0} = \omega_\alpha = 100$)  for ensembles of $N_A$ identical atoms as a function of time. 
\textbf{(c)} Time to $80\%$ saturation as a function of the number of atoms in the ensemble. We see the time monotonically decrease with atom number, as we expect due to the increased effective emission rate arising from the  collective superradiant effects. \textbf{(d)} Total occupation of the radiation modes normalized by the number of emitters in the ensemble, here for $N_A = 5$ atoms.} 
\label{fig:manyAtomRadiationOccupation}
\end{figure}

\begin{figure}[t] 
  \centering       \includegraphics[width=\columnwidth]{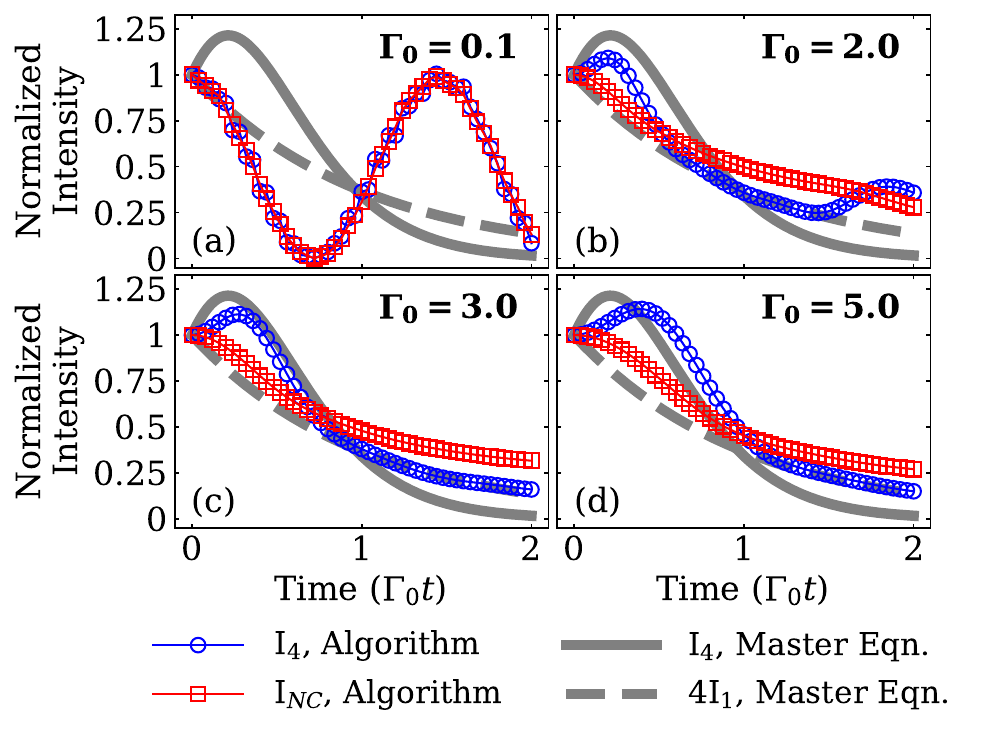}
      	\caption{Emission intensity for an ensemble of 4 atoms as a function of time. The gray lines in each panel correspond to the master equation solution for a homogeneous ensemble with the solid gray line representing the superradiant intensity and the dashed gray line representing the intensity for independent emission.
        The blue circles show the emission intensity $I_4(t)= \Gamma_0 \langle S^+S^-\rangle$
         calculated with the quantum algorithm. The red squares show the non-coherent contribution to the total emission intensity from the quantum algorithm $I_{4, \text{NC}}(t) = \Gamma_0\sum_\alpha\langle\sigma_\alpha^+\sigma_\alpha^-\rangle$.
         All curves are normalized by the emission intensity for a non-coherent ensemble of 4 emitters at $t=0$. 
         Each panel corresponds to a different value of the relaxation rate of individual atoms in the ensemble with $\Gamma_0 = 0.1, \;  2.0, \; 3.0, \; \mathrm{and} \; 5.0$. 
         } 
\label{fig:intensityInhomogeneousEnsemble}
\end{figure}

\subsection{Spectrally homogeneous ensembles}

\noindent We start by exploring the dynamics of a  spectrally homogeneous ensemble in the long wavelength limit where the atoms can be considered to be closely packed in space. Figure~\ref{fig:manyAtomRadiationOccupation} (a) presents the mode occupation as a function of time for a 5 atom system coupled to 7 radiation modes. This mode occupation corresponds to the emission spectrum of the system. Panel (b) shows the occupation of the central peak as a function of the number of atoms in the system. The darker shades in Panel (a) correspond to later times, while lighter shades in Panel (b) correspond to an increasing number of emitters in the ensemble. 

We observe through both of these graphs that the occupation number, even for the central/resonant mode never reaches the number of atoms. However, the total occupation number, across all the radiation modes, does saturate at long times when it reaches the number of atoms as shown in Panel (d) for $N_A=5$ emitters. Panel (c) shows the approach to saturation by plotting the time that the total mode occupation takes to reach 80\% of its saturation value as a function of the number of atoms. It clearly indicates that the time to saturation decreases with the number of atoms in the system, a hallmark of superradiant emission. Further benchmarking of the algorithm, including confirmation of the $N_A^2$ dependence of the maximum intensity on the number of emitters and an exploration of spatially diffuse ensembles can be found in Appendix~\ref{app:benchmarking}.

\begin{figure}[htbp] 
  \centering
        \includegraphics[width=\columnwidth]{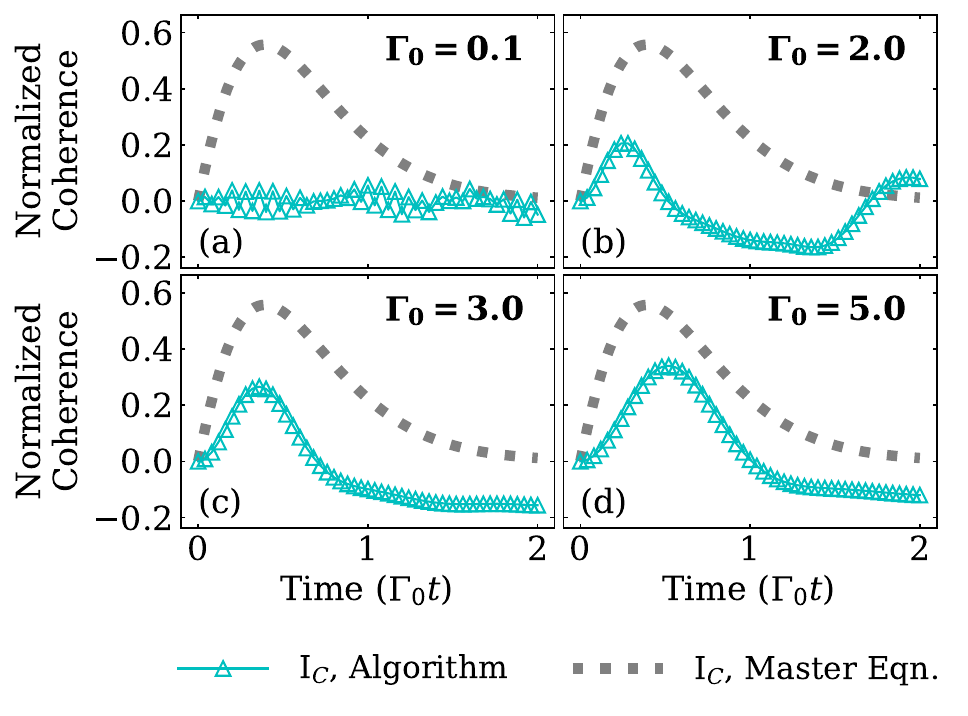}
      	\caption{Coherence $I_{4,C} = \Gamma_0\sum_{\alpha\neq\beta} \langle\sigma_\alpha^+ \sigma_\beta^-\rangle$ as a function of time for an ensemble of 4 atoms. The dotted gray line in each panel corresponds to the coherence of an ensemble of 4 perfectly resonant atoms that is calculated by numerically integrating the master equation using QuTiP. 
        The coherence of the diffuse ensemble simulated by the quantum algorithm is represented by the cyan triangles. As in Figure~\ref{fig:intensityInhomogeneousEnsemble}, all curves are normalized by $N_A\Gamma_0$, the emission rate of the ensemble at $t=0$ to allow for a direct comparison.}
	\label{fig:coherenceInhomogeneousEnsemble}
\end{figure}

\begin{figure}[t] 
  \centering
        \includegraphics[width=\columnwidth]{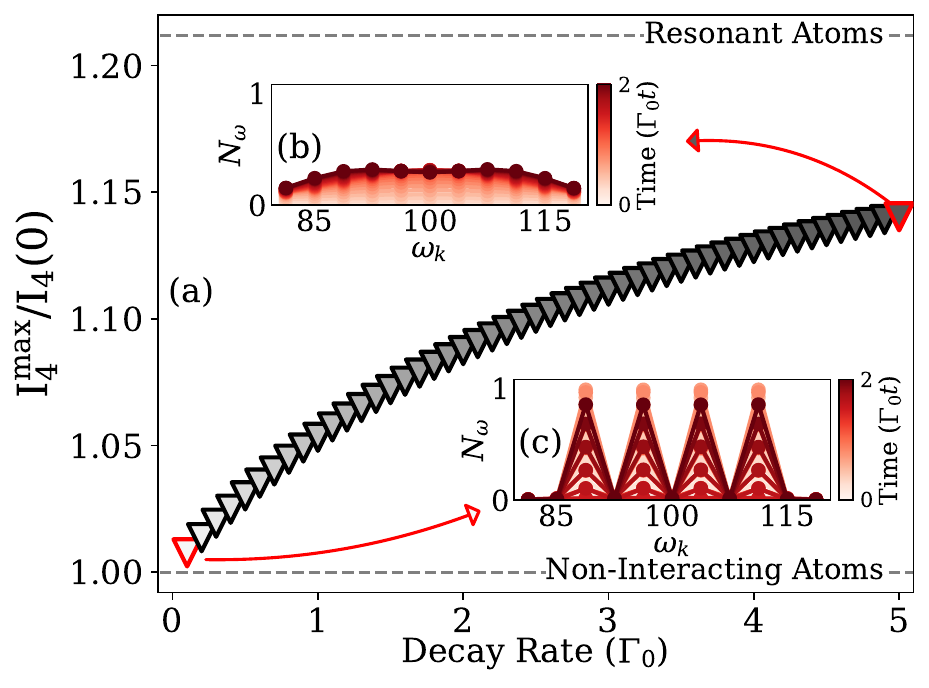}
      	\caption{\textbf{(a)} Normalized maximum intensity as a function of $\Gamma_0$ in the 4-atom system considered above. The color of the markers signify the effective overlap in atomic emission spectra, with darker shades corresponding to greater overlap. The dashed gray lines represent the expected maximum in intensity for the case of four resonant atoms (top) and four non-interacting atoms (bottom). \textbf{(b)} and \textbf{(c)} The emission spectrum (represented by the occupation number $N_\omega$ measured for the mode qubit at frequency $\omega$) as a function of time resulting from the decay in the same 4-atom system depicted respectively with $\Gamma_0 = 5$ and $\Gamma_0 = 0.1$ (highlighted in red at the ends of the curve in panel (a)).
        }
	\label{fig:spectraInhomogeneousEnsemble}
\end{figure}

\subsection{Spectrally inhomogeneous ensembles}
\label{subsec:inhomogeneousEnsembles}

\noindent It is particularly interesting to use the quantum circuits to explore how cooperative dynamics arises in a spectrally inhomogeneous ensemble where emitters in the ensemble have different emission frequencies despite being spatially dense. 
In what follows we consider the case of an ensemble of $N_A=4$ atoms in a bath of 11  bosonic modes. The bosonic modes are evenly distributed in frequency such that $\omega_k\in[\omega_{-5},\omega_5]$ have spacing $\Delta\omega_k = \omega_{i+1}-\omega_i = 3.75$. The atomic frequencies are distributed around the central mode frequency with $\Delta\omega_{\alpha} = 2\Delta\omega_k$ so that each atom's emission frequency is resonant with one of the modes. This frequency distribution is highlighted in Figure~\ref{fig:spectraInhomogeneousEnsemble}(c). Furthermore, we consider identical independent emission rates $\Gamma_0$ regardless of the natural emission frequencies of individual atoms.
Figure~\ref{fig:intensityInhomogeneousEnsemble} shows the emission intensity for this ensemble as a function of time. The gray lines in each panel correspond to the master equation solution for a homogeneous ensemble with the solid gray line representing the superradiant intensity and the dashed gray line representing the intensity for independent emission.
The blue circles show the emission intensity $I_4(t)= \Gamma_0 \langle S^+S^-\rangle$ calculated with the quantum algorithm. The red squares show the non-coherent contribution to the total emission intensity from the quantum algorithm $I_{4, \text{NC}}(t) = \Gamma_0\sum_\alpha\langle\sigma_\alpha^+\sigma_\alpha^-\rangle$.
All curves are normalized by their values at $t=0$, $N_A\Gamma_0$. 
When $\Gamma_0$ is small ($0.1$) we observe no photon enabled coupling between the atoms, each of which independently couples to the closest (resonant) radiation mode. Since the modes are sparsely distributed in comparison to the linewidth of individual emitters, this results in independent Rabi oscillations for each atom. For these small $\Gamma_0$ values, the total intensity overlaps with the non-coherent contribution.
As $\Gamma_0$ increases, we observe that the total emission rate $I_4(t)$ transitions from the non-coherent emission toward the perfectly resonant superradiant emission. 
The emergence of superradiance with increased free-emission linewidth or relaxation rate is confirmed by considering the coherence as a function of time for different $\Gamma_0$ values. This is captured by Figure~\ref{fig:coherenceInhomogeneousEnsemble} which shows the coherence $I_{4,C} = \Gamma_0\sum_{\alpha\neq\beta} \langle\sigma_\alpha^+ \sigma_\beta^-\rangle$ as a function of time. The dotted gray line in each panel corresponds to the coherence obtained through the integration of the master equation. 
The coherence of the diffuse ensemble simulated by the quantum algorithm is represented by the cyan triangles. As in Figure~\ref{fig:intensityInhomogeneousEnsemble}, all curves are normalized by $N_A\Gamma_0$, the emission rate of the ensemble at $t=0$, to allow for a direct comparison. We clearly see that increased $\Gamma_0$ leads to the emergence of a coherence peak corresponding to the superradiant burst identified in Figure~\ref{fig:intensityInhomogeneousEnsemble}. This phenomenon is highlighted in Figure~\ref{fig:spectraInhomogeneousEnsemble} which depicts the relationship between the ensemble's peak emission intensity and the linewidth of the individual emitters (panel (a)). The insets, panels (b) and (c) respectively, show the emission spectra for the overlapping individual emission spectra (large $\Gamma_0$) and the non-overlapping individual emission spectra (small $\Gamma_0$).

Note also in Figure~\ref{fig:spectraInhomogeneousEnsemble} Panels (b) and (c) that a single qubit allocation per radiation mode is sufficient even for the four-atom superradiance in this case, as no individual mode occupation increases above 1 at any time due to the spectral inhomogeneity of the ensemble. These spectra are also clearly affected by a low density of modes and limited spectral width. The former is the reason for which we see Rabi oscillations in the emission intensity in Figure~\ref{fig:intensityInhomogeneousEnsemble} (a) where the spectral overlap between emitters is small and each atom effectively interacts with only a single mode. As $\Gamma_0$ increases, this problem is abated and cooperative effects smoothly emerge in the ensemble, as depicted in the main panel (a) of this figure. 

\section{Conclusion}
\label{sec:Conclusion}

\noindent We adopt an efficient mapping of bosonic modes of radiation onto a set of quantum bits and implement quantum circuits for the simulation of an ensemble of many emitters coupled through a bath of radiation modes. Using these circuits, we are able to examine the emergence of cooperative dynamics, manifested by superradiant emission from the ensemble, as a function of various system parameters, including the number of emitters, the spectral inhomogeneity in the system, the emission lifetime of independent emitters, and the spatial separation between emitters. Since we avoid integrating out the bosonic bath and we explicitly track the dynamics of the entire system (emitters and bosonic modes of the radiation), our quantum circuits provide a clear and intuitive characterization of the many-emitter system and highlight subtle dependencies of cooperative dynamics on different system parameters. We are particularly able to characterize how superradiant emission can arise in an inhomogeneous ensemble as a function of the linewidth of the individual emitters.
The approach introduced in the present paper can be reliably applied to a broad set of light-matter interaction problems and has the capacity to shine new light on old problems, but also to serve as a powerful tool for immediate simulations of such problems. Furthermore, as the capacity of quantum processors improves, these methods can be readily adopted for systems that will likely remain untractable even with the most powerful classical computers.\\

\section*{Acknowledgment} 
\noindent We acknowledge support from the National Science Foundation under Grants No. PHY-2014023 and No. QIS-2328752. We thank T. Thomay and D. Schneble for useful discussions.

\bibliography{References}

@article{FotsoEtal_PRL2016,
  title = {Suppressing Spectral Diffusion of Emitted Photons with Optical Pulses},
  author = {Fotso, H. F. and Feiguin, A. E. and Awschalom, D. D. and Dobrovitski, V. V.},
  journal = {Phys. Rev. Lett.},
  volume = {116},
  issue = {3},
  pages = {033603},
  numpages = {6},
  year = {2016},
  month = {Jan},
  publisher = {American Physical Society},
  doi = {10.1103/PhysRevLett.116.033603},
  url = {https://link.aps.org/doi/10.1103/PhysRevLett.116.033603}
}

@article{Fotso_TPI_PRB_2019,
  title = {Pulse-enhanced two-photon interference with solid state quantum emitters},
  author = {Fotso, H. F. },
  journal = {Phys. Rev. B},
  volume = {100},
  issue = {9},
  pages = {094309},
  numpages = {6},
  year = {2019},
  month = {Sep},
  publisher = {American Physical Society},
  doi = {10.1103/PhysRevB.100.094309},
  url = {https://link.aps.org/doi/10.1103/PhysRevB.100.094309}
}

@book{Cohen_Tannoudji_Book1992,
  author = {C. Cohen-Tannoudji, C. and Dupont-Roc, J. and Grynberg, G.},
  year = {1992},
  title = {Atom-Photon Interactions, Basic Processes and Applications},
  publisher = {John Wiley \& Sons, Inc., New York}
}

@article{WeisskopfWigner1930,
author={Weisskopf, V.
and Wigner, E.},
title={Berechnung der nat{\"u}rlichen Linienbreite auf Grund der Diracschen Lichttheorie},
journal={Zeitschrift f{\"u}r Physik},
year={1930},
month={Jan},
day={01},
volume={63},
number={1},
pages={54-73},
issn={0044-3328},
doi={10.1007/BF01336768},
url={https://doi.org/10.1007/BF01336768}
}

@ARTICLE{JaynesCummings,
  author={Jaynes, E.T. and Cummings, F.W.},
  journal={Proceedings of the IEEE}, 
  title={Comparison of quantum and semiclassical radiation theories with application to the beam maser}, 
  year={1963},
  volume={51},
  number={1},
  pages={89-109},
  doi={10.1109/PROC.1963.1664}}

@misc{Qiskit2024,
      title={Quantum computing with {Q}iskit},
      author={Javadi-Abhari, Ali and Treinish, Matthew and Krsulich, Kevin and Wood, Christopher J. and Lishman, Jake and Gacon, Julien and Martiel, Simon and Nation, Paul D. and Bishop, Lev S. and Cross, Andrew W. and Johnson, Blake R. and Gambetta, Jay M.},
      year={2024},
      doi={10.48550/arXiv.2405.08810},
      eprint={2405.08810},
      archivePrefix={arXiv},
      primaryClass={quant-ph}
}

@book{Orszag2000quantum,
  title={Quantum Optics},
  author={Orszag, M.},
  isbn={9783540650089},
  lccn={99033296},
  series={Advanced texts in physics},
  url={https://books.google.com/books?id=v0YA4FYNk9QC},
  year={2000},
  publisher={Springer}
}

@misc{veis2016quantumchemistrybornoppenheimerapproximation,
      title={Quantum chemistry beyond Born-Oppenheimer approximation on a quantum computer: a simulated phase estimation study}, 
      author={Libor Veis and Jakub Višňák and Hiroaki Nishizawa and Hiromi Nakai and Jiří Pittner},
      year={2016},
      eprint={1507.03271},
      archivePrefix={arXiv},
      primaryClass={quant-ph},
      url={https://arxiv.org/abs/1507.03271}, 
}

@article{Kottmann_2021,
  doi = {10.1088/2058-9565/abfc94},
  url = {https://dx.doi.org/10.1088/2058-9565/abfc94},
  year = {2021},
  month = {aug},
  publisher = {IOP Publishing},
  volume = {6},
  number = {3},
  pages = {035010},
  author = {Kottmann, Jakob S and Krenn, Mario and Kyaw, Thi Ha and Alperin-Lea, Sumner and Aspuru-Guzik, Alán},
  title = {Quantum computer-aided design of quantum optics hardware},
  journal = {Quantum Science and Technology}
}

@article{GrossHaroche1982,
title = {Superradiance: An essay on the theory of collective spontaneous emission},
journal = {Physics Reports},
volume = {93},
number = {5},
pages = {301-396},
year = {1982},
issn = {0370-1573},
doi = {https://doi.org/10.1016/0370-1573(82)90102-8},
url = {https://www.sciencedirect.com/science/article/pii/0370157382901028},
author = {M. Gross and S. Haroche}
}

@article{Dicke1954,
  title = {Coherence in Spontaneous Radiation Processes},
  author = {Dicke, R. H.},
  journal = {Phys. Rev.},
  volume = {93},
  issue = {1},
  pages = {99--110},
  numpages = {0},
  year = {1954},
  month = {Jan},
  publisher = {American Physical Society},
  doi = {10.1103/PhysRev.93.99},
  url = {https://link.aps.org/doi/10.1103/PhysRev.93.99}
}

@article{BonifacioPRA1971_I,
  title = {Quantum Statistical Theory of Superradiance. II},
  author = {Bonifacio, R. and Schwendimann, P. and Haake, Fritz},
  journal = {Phys. Rev. A},
  volume = {4},
  issue = {3},
  pages = {854--864},
  numpages = {0},
  year = {1971},
  month = {Sep},
  publisher = {American Physical Society},
  doi = {10.1103/PhysRevA.4.854},
  url = {https://link.aps.org/doi/10.1103/PhysRevA.4.854}
}

@article{BonifacioPRA1971_II,
  title = {Quantum Statistical Theory of Superradiance. I},
  author = {Bonifacio, R. and Schwendimann, P. and Haake, Fritz},
  journal = {Phys. Rev. A},
  volume = {4},
  issue = {1},
  pages = {302--313},
  numpages = {0},
  year = {1971},
  month = {Jul},
  publisher = {American Physical Society},
  doi = {10.1103/PhysRevA.4.302},
  url = {https://link.aps.org/doi/10.1103/PhysRevA.4.302}
}

@misc{huang2022qubitizationbosons,
      title={Qubitization of Bosons}, 
      author={Xin-Yu Huang and Lang Yu and Xu Lu and Yin Yang and De-Sheng Li and Chun-Wang Wu and Wei Wu and Ping-Xing Chen},
      year={2022},
      eprint={2105.12563},
      archivePrefix={arXiv},
      primaryClass={quant-ph},
      url={https://arxiv.org/abs/2105.12563}, 
}

@misc{peng2023quantumsimulationbosonrelatedhamiltonians,
      title={Quantum Simulation of Boson-Related Hamiltonians: Techniques, Effective Hamiltonian Construction, and Error Analysis}, 
      author={Bo Peng and Yuan Su and Daniel Claudino and Karol Kowalski and Guang Hao Low and Martin Roetteler},
      year={2023},
      eprint={2307.06580},
      archivePrefix={arXiv},
      primaryClass={quant-ph},
      url={https://arxiv.org/abs/2307.06580}, 
}

@misc{escofet2024revisitingmappingquantumcircuits,
      title={Revisiting the Mapping of Quantum Circuits: Entering the Multi-Core Era}, 
      author={Pau Escofet and Anabel Ovide and Medina Bandic and Luise Prielinger and Hans van Someren and Sebastian Feld and Eduard Alarcón and Sergi Abadal and Carmen G. Almudéver},
      year={2024},
      eprint={2403.17205},
      archivePrefix={arXiv},
      primaryClass={quant-ph},
      url={https://arxiv.org/abs/2403.17205}, 
}

@article{jiang2025binary,
  title={A binary integer programming-based method for qubit mapping in sparse architectures},
  author={Jiang, Hui and Fu, Jianling and Deng, Yuxin and Wu, Jun},
  journal={Acta Informatica},
  volume={62},
  number={1},
  pages={1--23},
  year={2025},
  publisher={Springer}
}

@article{IppolitiKhemani_prx2021,
  title = {Many-Body Physics in the NISQ Era: Quantum Programming a Discrete Time Crystal},
  author = {Ippoliti, Matteo and Kechedzhi, Kostyantyn and Moessner, Roderich and Sondhi, S.L. and Khemani, Vedika},
  journal = {PRX Quantum},
  volume = {2},
  issue = {3},
  pages = {030346},
  numpages = {24},
  year = {2021},
  month = {Sep},
  publisher = {American Physical Society},
  doi = {10.1103/PRXQuantum.2.030346},
  url = {https://link.aps.org/doi/10.1103/PRXQuantum.2.030346}
}

@article{Preskill2018quantumcomputingin,
  doi = {10.22331/q-2018-08-06-79},
  url = {https://doi.org/10.22331/q-2018-08-06-79},
  title = {Quantum {C}omputing in the {NISQ} era and beyond},
  author = {Preskill, John},
  journal = {{Quantum}},
  issn = {2521-327X},
  publisher = {{Verein zur F{\"{o}}rderung des Open Access Publizierens in den Quantenwissenschaften}},
  volume = {2},
  pages = {79},
  month = aug,
  year = {2018}
}

@article{BhartiAspuru-Guzik_RMP2022,
  title = {Noisy intermediate-scale quantum algorithms},
  author = {Bharti, Kishor and Cervera-Lierta, Alba and Kyaw, Thi Ha and Haug, Tobias and Alperin-Lea, Sumner and Anand, Abhinav and Degroote, Matthias and Heimonen, Hermanni and Kottmann, Jakob S. and Menke, Tim and Mok, Wai-Keong and Sim, Sukin and Kwek, Leong-Chuan and Aspuru-Guzik, Al\'an},
  journal = {Rev. Mod. Phys.},
  volume = {94},
  issue = {1},
  pages = {015004},
  numpages = {69},
  year = {2022},
  month = {Feb},
  publisher = {American Physical Society},
  doi = {10.1103/RevModPhys.94.015004},
  url = {https://link.aps.org/doi/10.1103/RevModPhys.94.015004}
}

@article{DaleyZoller_Nature2022,
author={Daley, Andrew J.
and Bloch, Immanuel
and Kokail, Christian
and Flannigan, Stuart
and Pearson, Natalie
and Troyer, Matthias
and Zoller, Peter},
title={Practical quantum advantage in quantum simulation},
journal={Nature},
year={2022},
month={Jul},
day={01},
volume={607},
number={7920},
pages={667-676},
issn={1476-4687},
doi={10.1038/s41586-022-04940-6},
url={https://doi.org/10.1038/s41586-022-04940-6}
}

@article{Fauseweh2024,
author={Fauseweh, Benedikt},
title={Quantum many-body simulations on digital quantum computers: State-of-the-art and future challenges},
journal={Nature Communications},
year={2024},
month={Mar},
day={08},
volume={15},
number={1},
pages={2123},
issn={2041-1723},
doi={10.1038/s41467-024-46402-9},
url={https://doi.org/10.1038/s41467-024-46402-9}
}

@misc{qutip2025,
      title={QuTiP 5: The Quantum Toolbox in Python}, 
      author={Neill Lambert and Eric Giguère and Paul Menczel and Boxi Li and Patrick Hopf and Gerardo Suárez and Marc Gali and Jake Lishman and Rushiraj Gadhvi and Rochisha Agarwal and Asier Galicia and Nathan Shammah and Paul Nation and J. R. Johansson and Shahnawaz Ahmed and Simon Cross and Alexander Pitchford and Franco Nori},
      year={2025},
      eprint={2412.04705},
      archivePrefix={arXiv},
      primaryClass={quant-ph},
      url={https://arxiv.org/abs/2412.04705}, 
}

@article{RaimondEtAlPRL1982,
  title = {Collective Absorption of Blackbody Radiation by Rydberg Atoms in a Cavity: An Experiment on Bose Statistics and Brownian Motion},
  author = {Raimond, J. M. and Goy, P. and Gross, M. and Fabre, C. and Haroche, S.},
  journal = {Phys. Rev. Lett.},
  volume = {49},
  issue = {2},
  pages = {117--120},
  numpages = {0},
  year = {1982},
  month = {Jul},
  publisher = {American Physical Society},
  doi = {10.1103/PhysRevLett.49.117},
  url = {https://link.aps.org/doi/10.1103/PhysRevLett.49.117}
}

@article{SlamaEtAlPRL2007,
  title = {Superradiant Rayleigh Scattering and Collective Atomic Recoil Lasing in a Ring Cavity},
  author = {Slama, S. and Bux, S. and Krenz, G. and Zimmermann, C. and Courteille, Ph. W.},
  journal = {Phys. Rev. Lett.},
  volume = {98},
  issue = {5},
  pages = {053603},
  numpages = {4},
  year = {2007},
  month = {Feb},
  publisher = {American Physical Society},
  doi = {10.1103/PhysRevLett.98.053603},
  url = {https://link.aps.org/doi/10.1103/PhysRevLett.98.053603}
}

@article{Colombo2022,
author={Colombo, Simone
and Pedrozo-Pe{\~{n}}afiel, Edwin
and Adiyatullin, Albert F.
and Li, Zeyang
and Mendez, Enrique
and Shu, Chi
and Vuleti{\'{c}}, Vladan},
title={Time-reversal-based quantum metrology with many-body entangled states},
journal={Nature Physics},
year={2022},
month={Aug},
day={01},
volume={18},
number={8},
pages={925-930},
issn={1745-2481},
doi={10.1038/s41567-022-01653-5},
url={https://doi.org/10.1038/s41567-022-01653-5}
}

@article{Delanty_2011,
doi = {10.1088/1367-2630/13/5/053032},
url = {https://doi.org/10.1088/1367-2630/13/5/053032},
year = {2011},
month = {may},
publisher = {},
volume = {13},
number = {5},
pages = {053032},
author = {Delanty, M and Rebić, S and Twamley, J},
title = {Superradiance and phase multistability in circuit quantum electrodynamics},
journal = {New Journal of Physics}
}

@article{Franke2023,
author={Franke, Johannes
and Muleady, Sean R.
and Kaubruegger, Raphael
and Kranzl, Florian
and Blatt, Rainer
and Rey, Ana Maria
and Joshi, Manoj K.
and Roos, Christian F.},
title={Quantum-enhanced sensing on optical transitions through finite-range interactions},
journal={Nature},
year={2023},
month={Sep},
day={01},
volume={621},
number={7980},
pages={740-745},
issn={1476-4687},
doi={10.1038/s41586-023-06472-z},
url={https://doi.org/10.1038/s41586-023-06472-z}
}

@article{MasonEtAlPRL2020,
  title = {Many-Body Signatures of Collective Decay in Atomic Chains},
  author = {Masson, Stuart J. and Ferrier-Barbut, Igor and Orozco, Luis A. and Browaeys, Antoine and Asenjo-Garcia, Ana},
  journal = {Phys. Rev. Lett.},
  volume = {125},
  issue = {26},
  pages = {263601},
  numpages = {7},
  year = {2020},
  month = {Dec},
  publisher = {American Physical Society},
  doi = {10.1103/PhysRevLett.125.263601},
  url = {https://link.aps.org/doi/10.1103/PhysRevLett.125.263601}
}

@article{ClemensEtAl2003,
  title = {Collective spontaneous emission from a line of atoms},
  author = {Clemens, J. P. and Horvath, L. and Sanders, B. C. and Carmichael, H. J.},
  journal = {Phys. Rev. A},
  volume = {68},
  issue = {2},
  pages = {023809},
  numpages = {19},
  year = {2003},
  month = {Aug},
  publisher = {American Physical Society},
  doi = {10.1103/PhysRevA.68.023809},
  url = {https://link.aps.org/doi/10.1103/PhysRevA.68.023809}
}

@article{Brinke2015,
  title = {Dicke superradiance as a nondestructive probe for quantum quenches in optical lattices},
  author = {ten Brinke, Nicolai and Sch\"utzhold, Ralf},
  journal = {Phys. Rev. A},
  volume = {92},
  issue = {1},
  pages = {013617},
  numpages = {5},
  year = {2015},
  month = {Jul},
  publisher = {American Physical Society},
  doi = {10.1103/PhysRevA.92.013617},
  url = {https://link.aps.org/doi/10.1103/PhysRevA.92.013617}
}

@article{DeMille2024,
author={DeMille, David
and Hutzler, Nicholas R.
and Rey, Ana Maria
and Zelevinsky, Tanya},
title={Quantum sensing and metrology for fundamental physics with molecules},
journal={Nature Physics},
year={2024},
month={May},
day={01},
volume={20},
number={5},
pages={741-749},
issn={1745-2481},
doi={10.1038/s41567-024-02499-9},
url={https://doi.org/10.1038/s41567-024-02499-9}
}

@inproceedings{FotsoSPIE2025,
author = {Herbert F. Fotso and Shreekanth S. Yuvarajan and Vincent Iglesias-Cardinale and David Hucul},
title = {{Quantum computation approaches for modeling photon-mediated operations in quantum information processing}},
volume = {13604},
booktitle = {Optics and Photonics for Information Processing XIX},
editor = {Khan M. Iftekharuddin and Abdul A. S. Awwal and Victor Hugo Diaz-Ramirez and Andr{\'e}s M{\'a}rquez},
organization = {International Society for Optics and Photonics},
publisher = {SPIE},
pages = {136040K},
year = {2025},
doi = {10.1117/12.3065016},
URL = {https://doi.org/10.1117/12.3065016}
}

@article{Fotso_noisyTLS2022,
  title = {Tuning spectral properties of individual and multiple quantum emitters in noisy environments},
  author = {Fotso, Herbert F.},
  journal = {Phys. Rev. A},
  volume = {107},
  issue = {2},
  pages = {023719},
  numpages = {7},
  year = {2023},
  month = {Feb},
  publisher = {American Physical Society},
  doi = {10.1103/PhysRevA.107.023719},
  url = {https://link.aps.org/doi/10.1103/PhysRevA.107.023719}
}

@article{KoppenhoeferPRL2023,
  title = {Squeezed Superradiance Enables Robust Entanglement-Enhanced Metrology Even with Highly Imperfect Readout},
  author = {Koppenh\"ofer, Martin and Groszkowski, Peter and Clerk, A. A.},
  journal = {Phys. Rev. Lett.},
  volume = {131},
  issue = {6},
  pages = {060802},
  numpages = {7},
  year = {2023},
  month = {Aug},
  publisher = {American Physical Society},
  doi = {10.1103/PhysRevLett.131.060802},
  url = {https://link.aps.org/doi/10.1103/PhysRevLett.131.060802}
}

@article{BaskicPRL2014,
  title = {Controlling Discrete and Continuous Symmetries in ``Superradiant'' Phase Transitions with Circuit QED Systems},
  author = {Baksic, Alexandre and Ciuti, Cristiano},
  journal = {Phys. Rev. Lett.},
  volume = {112},
  issue = {17},
  pages = {173601},
  numpages = {5},
  year = {2014},
  month = {Apr},
  publisher = {American Physical Society},
  doi = {10.1103/PhysRevLett.112.173601},
  url = {https://link.aps.org/doi/10.1103/PhysRevLett.112.173601}
}

@article{Lei2023,
author={Lei, Mi
and Fukumori, Rikuto
and Rochman, Jake
and Zhu, Bihui
and Endres, Manuel
and Choi, Joonhee
and Faraon, Andrei},
title={Many-body cavity quantum electrodynamics with driven inhomogeneous emitters},
journal={Nature},
year={2023},
month={May},
day={01},
volume={617},
number={7960},
pages={271-276},
issn={1476-4687},
doi={10.1038/s41586-023-05884-1},
url={https://doi.org/10.1038/s41586-023-05884-1}
}

@article{ALPS2,
doi = {10.1088/1742-5468/2011/05/P05001},
url = {https://doi.org/10.1088/1742-5468/2011/05/P05001},
year = {2011},
month = {may},
publisher = {},
volume = {2011},
number = {05},
pages = {P05001},
author = {Bauer, B and Carr, L D and Evertz, H G and Feiguin, A and Freire, J and Fuchs, S and Gamper, L and Gukelberger, J and Gull, E and Guertler, S and Hehn, A and Igarashi, R and Isakov, S V and Koop, D and Ma, P N and Mates, P and Matsuo, H and Parcollet, O and Pawłowski, G and Picon, J D and Pollet, L and Santos, E and Scarola, V W and Schollwöck, U and Silva, C and Surer, B and Todo, S and Trebst, S and Troyer, M and Wall, M L and Werner, P and Wessel, S},
title = {The ALPS project release 2.0: open source software for strongly correlated systems},
journal = {Journal of Statistical Mechanics: Theory and Experiment}
}

@article{TRIQS,
title = {TRIQS: A toolbox for research on interacting quantum systems},
journal = {Computer Physics Communications},
volume = {196},
pages = {398-415},
year = {2015},
issn = {0010-4655},
doi = {https://doi.org/10.1016/j.cpc.2015.04.023},
url = {https://www.sciencedirect.com/science/article/pii/S0010465515001666},
author = {Olivier Parcollet and Michel Ferrero and Thomas Ayral and Hartmut Hafermann and Igor Krivenko and Laura Messio and Priyanka Seth}
}

@article{QE-2017,
  author={P Giannozzi and O Andreussi and T Brumme and O Bunau and M Buongiorno Nardelli
  and M Calandra and R Car and C Cavazzoni and D Ceresoli and M Cococcioni and N Colonna
  and I Carnimeo and A Dal Corso and S de Gironcoli and P Delugas and R A DiStasio Jr and A Ferretti
  and A Floris and G Fratesi and G Fugallo and R Gebauer and U Gerstmann and F Giustino and T Gorni
  and J Jia and M Kawamura and H-Y Ko and A Kokalj and E Küçükbenli and M Lazzeri and M Marsili
  and N Marzari and F Mauri and N L Nguyen and H-V Nguyen and A Otero-de-la-Roza and L Paulatto
  and S Poncé and D Rocca and R Sabatini and B Santra and M Schlipf and A P Seitsonen
  and A Smogunov and I Timrov and T Thonhauser and P Umari and N Vast and X Wu and S Baroni},
  title={Advanced capabilities for materials modelling with QUANTUM ESPRESSO},
  journal={Journal of Physics: Condensed Matter},
  volume={29},
  number={46},
  pages={465901},
  url={http://stacks.iop.org/0953-8984/29/i=46/a=465901},
  year={2017},
}

@misc{Ansys,
  title={Ansys Simulation Software},
  url={{https://www.ansys.com/.}},
}

@misc{COMSOL,
  title={COMSOL - Software for Multiphysics Simulation},
  url={{https://www.comsol.com/.}},
}

@article{HosseinabadiOksanaMarinoPRXquantum2025,
  title = {User-Friendly Truncated Wigner Approximation for Dissipative Spin Dynamics},
  author = {Hosseinabadi, Hossein and Chelpanova, Oksana and Marino, Jamir},
  journal = {PRX Quantum},
  volume = {6},
  issue = {3},
  pages = {030344},
  numpages = {23},
  year = {2025},
  month = {Sep},
  publisher = {American Physical Society},
  doi = {10.1103/1wwv-k7hg},
  url = {https://link.aps.org/doi/10.1103/1wwv-k7hg}
}

@article{MotAsenjo-GarciaEtAlPRL2023,
  title = {Dicke Superradiance Requires Interactions beyond Nearest Neighbors},
  author = {Mok, Wai-Keong and Asenjo-Garcia, Ana and Sum, Tze Chien and Kwek, Leong-Chuan},
  journal = {Phys. Rev. Lett.},
  volume = {130},
  issue = {21},
  pages = {213605},
  numpages = {7},
  year = {2023},
  month = {May},
  publisher = {American Physical Society},
  doi = {10.1103/PhysRevLett.130.213605},
  url = {https://link.aps.org/doi/10.1103/PhysRevLett.130.213605}
}

@article{HP_Transform,
  title = {Field Dependence of the Intrinsic Domain Magnetization of a Ferromagnet},
  author = {Holstein, T. and Primakoff, H.},
  journal = {Phys. Rev.},
  volume = {58},
  issue = {12},
  pages = {1098--1113},
  numpages = {0},
  year = {1940},
  month = {Dec},
  publisher = {American Physical Society},
  doi = {10.1103/PhysRev.58.1098},
  url = {https://link.aps.org/doi/10.1103/PhysRev.58.1098}
}

@article{Masson_2020_AWGQED,
   title={Atomic-waveguide quantum electrodynamics},
   volume={2},
   ISSN={2643-1564},
   url={http://dx.doi.org/10.1103/PhysRevResearch.2.043213},
   DOI={10.1103/physrevresearch.2.043213},
   number={4},
   journal={Physical Review Research},
   publisher={American Physical Society (APS)},
   author={Masson, Stuart J. and Asenjo-Garcia, Ana},
   year={2020},
   month=nov }

@article{Shen:05,
author = {J. T. Shen and Shanhui Fan},
journal = {Opt. Lett.},
number = {15},
pages = {2001--2003},
publisher = {Optica Publishing Group},
title = {Coherent photon transport from spontaneous emission in one-dimensional waveguides},
volume = {30},
month = {Aug},
year = {2005},
url = {https://opg.optica.org/ol/abstract.cfm?URI=ol-30-15-2001},
doi = {10.1364/OL.30.002001},
}

@article{Blais2004,
  title = {Cavity quantum electrodynamics for superconducting electrical circuits: An architecture for quantum computation},
  author = {Blais, Alexandre and Huang, Ren-Shou and Wallraff, Andreas and Girvin, S. M. and Schoelkopf, R. J.},
  journal = {Phys. Rev. A},
  volume = {69},
  issue = {6},
  pages = {062320},
  numpages = {14},
  year = {2004},
  month = {Jun},
  publisher = {American Physical Society},
  doi = {10.1103/PhysRevA.69.062320},
  url = {https://link.aps.org/doi/10.1103/PhysRevA.69.062320}
}

@misc{yuvarajan2025,
      title={Cavity Mediated Two-Qubit Gate: Tuning to Optimal Performance with NISQ Era Quantum Simulations}, 
      author={Shreekanth S. Yuvarajan and Vincent Iglesias-Cardinale and David Hucul and Herbert F. Fotso},
      year={2025},
      eprint={2512.12030},
      archivePrefix={arXiv},
      primaryClass={quant-ph},
      url={https://arxiv.org/abs/2512.12030}, 
}

@article{Glauber1963,
  title = {The Quantum Theory of Optical Coherence},
  author = {Glauber, Roy J.},
  journal = {Phys. Rev.},
  volume = {130},
  issue = {6},
  pages = {2529--2539},
  numpages = {0},
  year = {1963},
  month = {Jun},
  publisher = {American Physical Society},
  doi = {10.1103/PhysRev.130.2529},
  url = {https://link.aps.org/doi/10.1103/PhysRev.130.2529}
}

\newpage
\onecolumngrid
\appendix

\newpage
\section{Benchmarking Algorithm with Spectrally Homogeneous Ensemble Results}
\label{app:benchmarking}

We present here some standard results in the study of superradiance, as reproduced by the proposed quantum algorithm.

\subsection{Spatially Identical Atoms}

\begin{figure}[h] 
  \centering
      	\includegraphics[width=18.0cm]{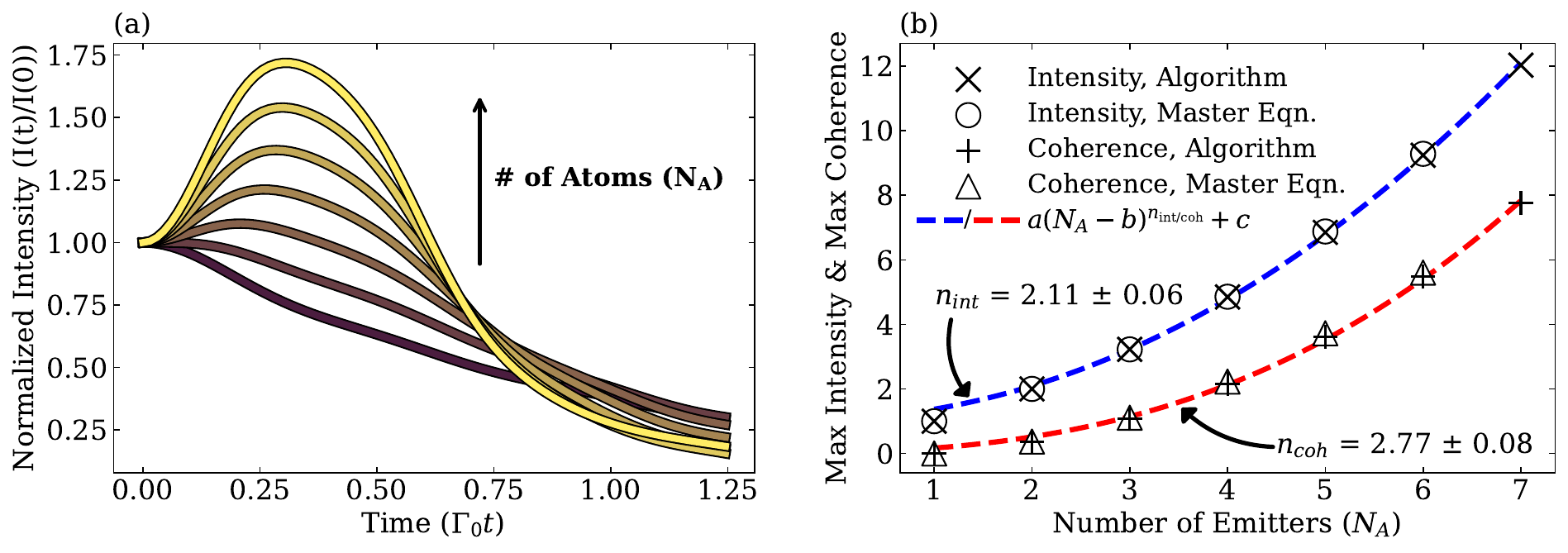}
      	\caption{\textbf{(a)} Intensity as a function of time for ensembles of $N_A=1$ (dark brown line) to $N_A=7$ (yellow line) emitters. All curves are normalized by $N_A\Gamma_0$, their value at $t=0$, for purposes of direct comparison. \textbf{(b)} Maximum intensity (circles and x's) and maximum coherence (triangles and pluses) achieved by spectrally homogeneous ensembles of $N_A=1$ to $7$ atoms in a bath of seven bosonic modes. Each set of data is fit with the curve $f_{\text{fit}} = a(N_A-b)^n+c$ and the fits are plotted in red for the coherence and blue for the intensity with the exponent of best fit labeled accordingly.}
        \label{fig:benchmarking}
\end{figure}

First, we compute the intensity of the radiation, defined in Eq.(\ref{eq:intensity1}), for a homogeneous, dense ensemble of $N_A\in[1,7]$ atoms in the long wavelength limit and a bath of $N_M=7$ radiation modes.  Figure~\ref{fig:benchmarking} (a) presents this intensity as a function of time for different numbers $N_A$ of atoms in the ensemble. Lighter shades indicate larger numbers of atoms. The intensity is normalized to its value at time $t = 0$. The ensemble is coupled to 7 radiation modes that are evenly distributed in frequency with $\omega_k\in[-25, 25]$. The figure features the emergence of the superradiant burst and, when the number of atoms is increased, the characteristic peak amplitude increases with the number of atoms. Note the faster decay in the intensity for larger $N_A$, but we also predictably see faster re-excitation of the atoms as $N_A$ increases.

The specific dependence of this superradiant peak amplitude on the number of atoms is captured in Fig.~\ref{fig:benchmarking} (b). The figure also shows the maximum coherence, defined by Eq.(\ref{eq:coherence}), as a function of the number of atoms in the ensemble. Both the intensity and coherence in this figure are normalized by $\Gamma_0$. We see a satisfactory agreement with the results from the QuTiP simulation of the master equation for the same number of atoms, despite the fact that the master equation assumes coupling of the atomic ensemble to an infinite number of bosonic modes. This highlights the performance of the quantum algorithm even with the reduced number of bosonic modes of radiation. The lines indicate a polynomial fit to the calculated data, with a slight deviation from the expected quadratic dependence of the intensity peak on the number of atoms that can be ascribed to the limited system size.\\ 

\subsection{Spatially Diffuse Atoms}
\begin{figure}[h] 
  \centering
      	\includegraphics[width=18.0cm]{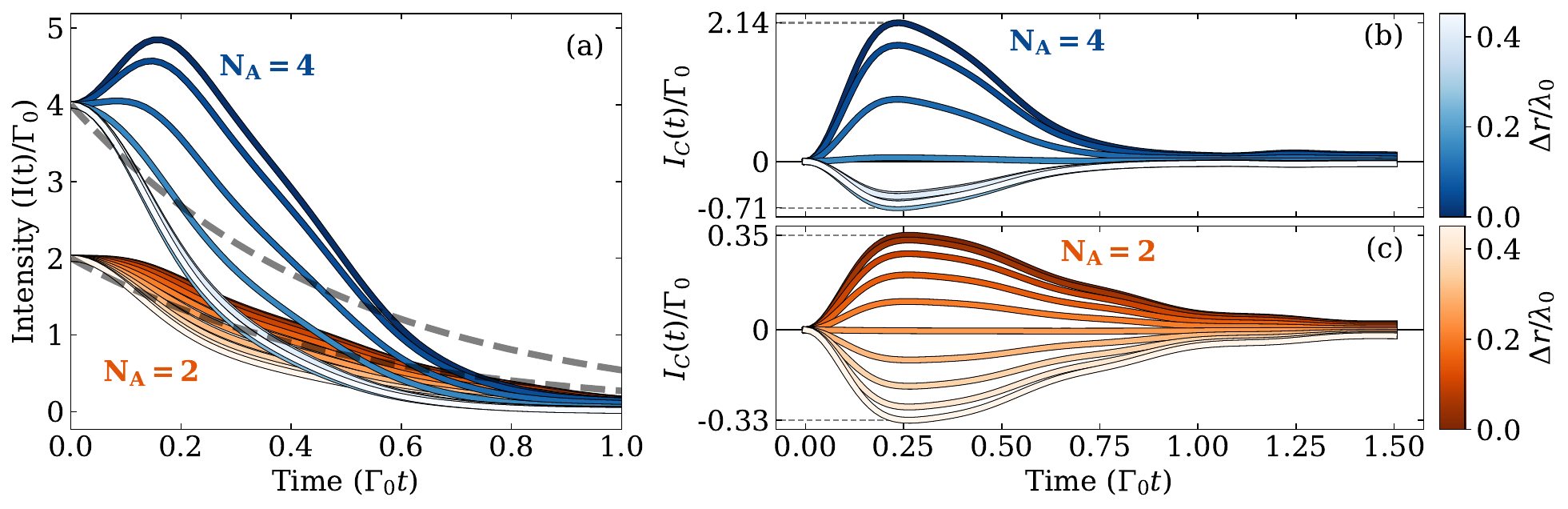}
      	\caption{\textbf{(a)} Intensity as a function of time for spectrally homogeneous ensembles of $N_A=2$ and $N_A=4$ atoms in orange and blue, respectively. In both cases, the intensities exhibit sub-exponential decays even when inter-emitter separation is significant on the scale of emission wavelength. This is to be expected in a one dimensional waveguide, as discussed in the text. \textbf{(b)} and \textbf{(c)} Coherence as a function of time for spectrally homogeneous ensembles of $N_A=2$ and $N_A=4$ atoms in orange and blue, respectively.}
        \label{fig:SpatiallyDilute4Atoms}
\end{figure}

Finally, we study the emergence of cooperative emission as a function of the spatial density of the ensemble. Up until this point, we have effectively treated all wavevectors $\mathbf k$ as indices of the radiation modes. More care must be given to their treatment when considering interaction between the radiation bath and a physical configuration of the atoms in space. We use natural units ($c\equiv 1$) so that $|\mathbf k| = k = \omega_{\mathbf k}$ and consider a one dimensional vacuum with atoms evenly distributed along the spatial axis with inter-atomic separation $\Delta r$.
As seen in Fig.~\ref{fig:benchmarking}, the long wavelength limit of this system
(i.e $r/\lambda_0 = kr <<1 \longrightarrow r<<1/\omega_{\mathbf k\text{,max}}$)
shows a clear superradiant peak for $N_A>2$. In Fig.~\ref{fig:SpatiallyDilute4Atoms}, we plot the intensity as a function of time for different values of the spatial separation $\Delta r/\lambda_0$ between consecutive emitters in a linear configuration of length 2 (orange) and 4 (blue) measured in units of the emission wavelength. Darker coloring in the figure corresponds to smaller inter-atomic spacing.
The superradiant peak clearly emerges when $\Delta r/\lambda_0$ is sufficiently small.
but the intensity profiles exhibit a sub-exponential decay for larger inter-atom spacings. This behavior is expected when photon propagation is confined to a single dimension~\cite{Masson_2020_AWGQED, Shen:05, Blais2004}.

To understand this behavior, consider the generalized definition of the intensity given in Eq.~\ref{eq:intensity1}~\cite{Orszag2000quantum,Glauber1963},

\begin{equation}
    I(t) = \sum_{\alpha\beta}\Gamma_{\alpha\beta}\langle\sigma_\alpha^+\sigma_\beta^-\rangle \propto \langle E^-(t;\mathbf r)E^+(t;\mathbf r)\rangle,
    \label{eq:intensity4}
\end{equation}

\noindent where now the effective emission rate depends on photon induced coupling between atoms $\alpha$ and $\beta$. To find a form for this emission rate we examine, without specifying dimensionality at this stage, the electric field induced by photon emission from atom $\alpha$,  $\mathbf E_\alpha = \mathbf E_{\alpha,0} + \mathbf E_{\alpha,RR}$. The radiation reaction term is given by

\begin{equation}
    \mathbf E_{\alpha, RR}^+(t, \mathbf r_\alpha) =-i\sum_\beta\sum_{\mathbf k\lambda}\sqrt{\frac{\hbar\omega_{\mathbf k}}{2\epsilon_0 v}}g_{\mathbf k\lambda}\mathbf e_{\mathbf k\lambda}e^{i\mathbf k\cdot(\mathbf r_\alpha-\mathbf r_\beta)}\int_0^tdt'\sigma_\beta^-(t')e^{i\omega_{\mathbf k}(t'-t)},
    \label{eqn:RadiationReaction}
\end{equation}

\noindent from which the definition of $\Gamma_{\alpha\beta}$ is readily found. Restricting now our system to a single dimension of photon propagation, we make the approximation $\sum_{\mathbf k\lambda}\rightarrow\frac{L}{2\pi}\int_0^\infty dk$ which, after inserting $g_{\mathbf k\lambda}=g_k=\sqrt{\frac{\omega_k}{2\epsilon_0L\hbar}}\mathbf d$, transforms Eq.~\ref{eqn:RadiationReaction} into

\begin{equation}
    \mathbf E_{\alpha, RR}^+(t, \mathbf r_\alpha) =-i\frac{d}{4\pi\epsilon_0c}\sum_\beta\int_0^t dt'\sigma_\beta^-(t')\int_0^\infty d\omega_k\omega_ke^{i\omega_k(t'-t)}e^{i\omega_kr_{\alpha\beta}/c}
    \label{eqn:RadiationReaction2}
\end{equation}

\noindent where we have set the dipole matrix element $d=\mathbf d\cdot\mathbf e_\mathbf k$ for all atoms in the ensemble. Using $\sigma_\beta^-(t')=\sigma_\beta(t)e^{i\omega_0(t-t')}$, in which $\hbar\omega_0=\hbar\omega_\beta$ is the energy gap between ground and excited states for all atoms, and the Markov approximation ($\int_0^tdt'e^{-ix(t-t')}\rightarrow\pi\delta (x)$) we immediately find

\begin{equation}
    \begin{aligned}
    \mathbf E_{\alpha, RR}^+(t, \mathbf r_\alpha) &=-\frac{id\omega_0}{4\epsilon_0c}\sum_\beta\sigma_\beta^-(t)e^{i\omega_0r_{\alpha\beta}/c}\\
    &=-\frac{id\omega_0}{4\epsilon_0c}\sum_\beta\sigma_\beta^-(t)e^{ir_{\alpha\beta}/\lambda_0}\\
    &=-i\frac{\hbar}{d}\Gamma_0\sum_\beta\Gamma_{\alpha\beta}'\sigma_\beta^-(t).
    \end{aligned}
    \label{eqn:RadiationReaction3}
\end{equation}

In deriving Eq.~\ref{eqn:RadiationReaction3}, we have assumed all atoms in the one dimensional ensemble share an emission wavelength of $\lambda_0=c/\omega_0$ and thus define $\Gamma_0 = \frac{d^2\omega_0^2}{4\epsilon_0c\hbar}$ and $\Gamma_{\alpha\beta}'=e^{ir_{\alpha\beta}/\lambda_0}$. We can then write $\Gamma_{\alpha\beta} = \Gamma_0\Gamma_{\alpha\beta}'$ which, for our one dimensional ensemble, is clearly periodic in $r_{\alpha\beta}/\lambda_0$ and approaches $\Gamma_0$ as $r_{\alpha\beta}$ approaches 0.

We can verify both this form of the collective emission rate and the quantum algorithm by affirming that the latter reproduces results predicted by the former. We rewrite the coherence (the intensity given by Eq.~\ref{eq:intensity4} with the $\alpha-\alpha$ terms subtracted away) by wrapping the time dependence into coefficients $A_{\alpha\beta}(t;r_1,r_2,\cdots,r_{N_A}) = \langle\sigma_\alpha^+(t)\sigma_\beta^-(t)\rangle$ and noticing the symmetries inherent in the system, $r_{\alpha\beta} = -r_{\beta\alpha}$ and $A_{\alpha\beta} = A_{\beta\alpha}$,

\begin{equation}
    I_C(t) = 2\Gamma_0\sum_{\alpha<\beta}A_{\alpha\beta}(t)\cos(r_{\alpha\beta}/\lambda_0)
    \label{eqn:coherence}
\end{equation}

If we further specify our system so that $r_{12} = r_{23} = r_{34}=\cdots=\Delta r$ and define $A_i=A_{\alpha\beta}$ for $i=|\alpha-\beta|$ we can find generally

\begin{equation}
    I_C(t) = 2\Gamma_0\sum_{n=1}^{N_A}(N_A-n)A_{n}\cos(n\Delta r/\lambda_0)
    \label{eqn:coherence2}
\end{equation}

\begin{figure}[h]
  \centering
      	\includegraphics[width=7in]{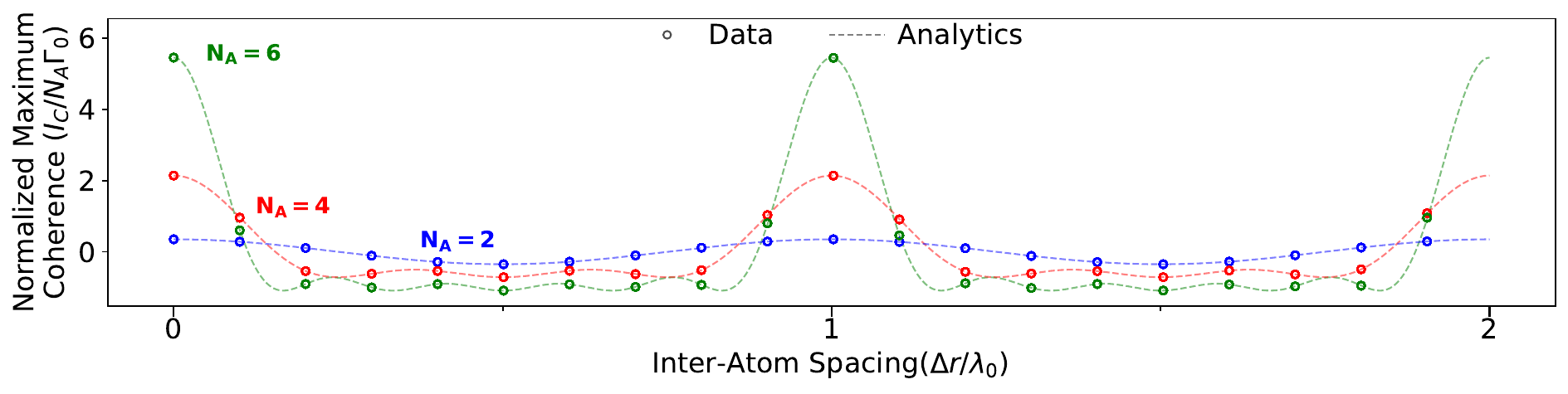}
      	\caption{Maximum coherence, which occurs always at the same time regardless of interatomic spacing (see Figure~\ref{fig:SpatiallyDilute4Atoms} panels (b) and (c)) as a function of interatomic distance for $N_A=2$, $N_A=4$ and $N_A=6$ atoms in blue, red, and green, respectively.}
	\label{fig:1dvacuum}
\end{figure}

What is left is to find an explicit form for the expectation values, $A_i$, which in general would require a mean field or master equation approach. Instead, we note that in the limit $\Delta r\rightarrow0$ and at the time of maximum coherence they must be constants that satisfy the condition $A_n = \max(I_C)\bigg/2\Gamma_0\sum_{n=1}^{N_A}(N_A-n)$ for all $n$. By plotting the maximum coherence as a function of interatomic spacing in a one dimensional array where the emitted radiation is confined to a single dimension as measured by the quantum algorithm and, separately, predicted by Eq.~\ref{eqn:coherence2}, we find near perfect agreement between the two solutions. This is depicted in Figure~\ref{fig:1dvacuum}.

\newpage
\section{Quantum Hardware Trials}
\label{app:hardware}

\noindent Although the quantum algorithms developed here are compatible with, and may be run on, NISQ era quantum systems, the circuit depths (number of gates) required after Trotterization remain a significant hurdle. To validate the algorithm as depicted in Figure~\ref{fig:circuit}, we build the circuits for one atom with emission frequency $\omega_\alpha=100$ and natural emission rate $\Gamma_0=2$ coupled to a bath of seven modes with evenly spaced frequencies $\omega_k\in[85, 115]$ with a single qubit allocation for each mode. The time axis is discretized into 8 Trotter steps in the time range $t\in[0,3/\Gamma_0]$.

The circuit is then run on three platforms:

\begin{enumerate}[label=(\alph*), leftmargin=4em]
    \item IBM Torino, a true quantum computer,
    \item a classical-quantum (noisy) simulator which reads the one- and two-qubit gate fidelities reported by a chosen backend (IBM Torino in this case) and attributes those fidelities to each gate in the simulation, and
    \item an ideal simulator which assumes 100\% fidelity for all gates.
\end{enumerate}

\begin{figure}[h] 
  \centering
      	\includegraphics[width=10.0cm]{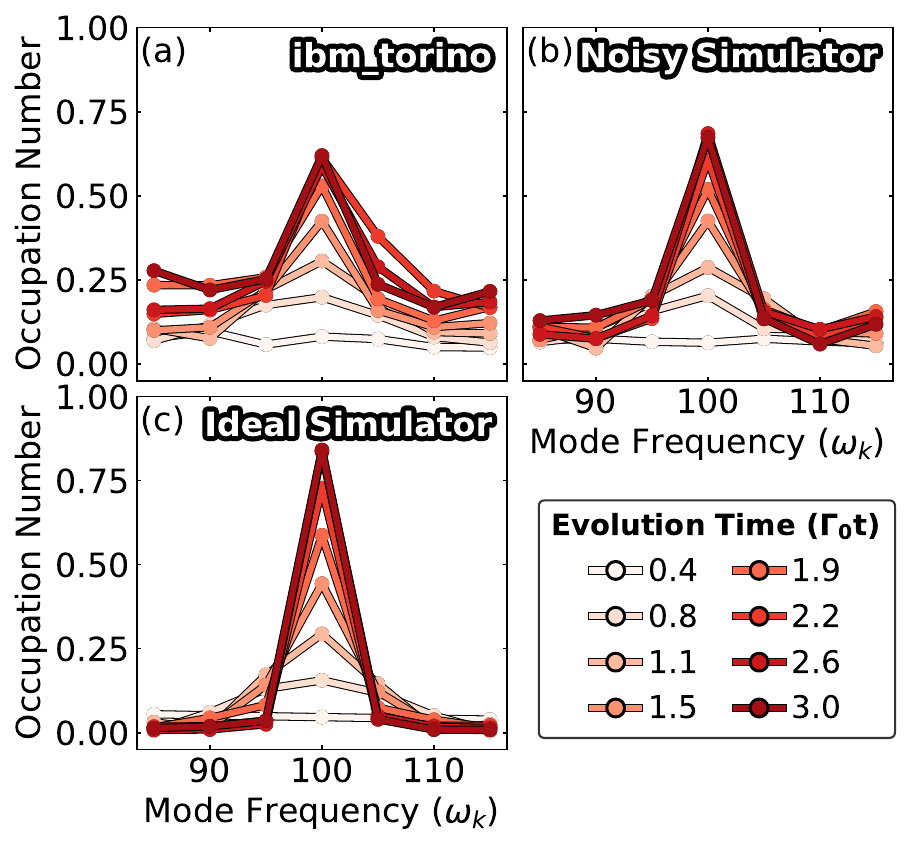}
      	\caption{Simulation results when run on (a) IBM Torino, (b) a noisy simulator, (c) an ideal simulator.}
        \label{fig:real}
\end{figure}

Due to atom-field coupling, we expect to see atomic decay resulting in a Lorentzian emission spectrum. The spectra resulting from each simulation are presented in Figure~\ref{fig:real}, represented by the occupation of each mode as a function of time.

\noindent Figure~\ref{fig:real} clearly depicts the mode resonant with the atom achieving the greatest occupation regardless of simulation type, however noise results in greater occupation of off-resonant modes on the quantum hardware and noisy simulator. On quantum hardware, this noise is largely due to thermalization of the system's qubits over the computation time and suboptimal gate fidelities. As the simulated system grows (i.e. as the number of atoms, modes, and/or the number of Trotter steps increase), the number of qubits, gate depth, and computation time grow as well. For this reason, we use the ideal simulator for all results presented in the main text where the number of atoms, the number of modes and especially the number of Trotter steps far exceed this toy model. As quantum hardware rapidly improves, these algorithms will likely be implemented to far greater effect than even the ideal simulator can achieve.


\newpage
\section{Proof of Recursion Relation for Creation operator in the Binary Representation}
\label{app: recursion relation}

\noindent We define the parameterized creation operator in the binary representation for bosonic mode $k$ with $q_k$ qubits allocated to its representation:

\begin{equation}
    a_{q_k}^\dagger(c) = \sum_{n=0}^{2^{q_k}-2}\sqrt{c+(n+1)}|n+1\rangle_{q_k q_k}\langle n|.
    \label{eqn:adag_of_c}
\end{equation}

The upper bound on the sum in Eq.~\ref{eqn:adag_of_c} arises due to the fact that the maximum number of photons in a mode modeled with $q_k$ qubits is $2^{q_k}-1$. The parameter $c$ represents an addition to the number of photons counted in mode $k$ when the number operator $n_{q_k}=a_{q_k}^\dagger a_{q_k}$ acts on that mode's state ket, $|n\rangle_{q_k}$. It exists only as a convenient notation to be used when writing the recursion relation that constructs the bosonic operators and should be set to zero to obtain physical results in any simulation.

The state ket in Eq.~\ref{eqn:adag_of_c} takes the form

\begin{equation}
    |n\rangle_{q_k} = |\underbrace{00101\cdots0111}_{\mathclap{\text{bit-string (length $q_k$) representation of integer $n$}}}\rangle
\end{equation}

For example, $|5\rangle_3$ will be represented as $|5\rangle_3=|1\rangle\otimes|0\rangle\otimes|1\rangle = |101\rangle$, where we omit the subscript indicating bit-string length when the length is one.

From Eq.~\ref{eqn:adag_of_c} we can find the parameterized creation operator for the lowest possible qubit allocation,

\begin{equation}
    a_1^\dagger(c) = \sqrt{c+1}|1\rangle\langle0|.
    \label{eqn: adag1 of c}
\end{equation}

Clearly a qubit allocation lower than 1 is not physical. For this reason, we define $a_0^\dagger(c) = 0$.

Additionally, we will define the exponentiation of the tensor product $A^{\otimes \xi} = \underbrace{A\otimes A\otimes\cdots\otimes A}_{\text{$\xi$ many times}}$. 
We define $A^{\otimes 0} = 1$.

With these definitions, we will now prove by induction the following recursion relation for $a_{q_k}^\dagger(c)$ :

\begin{equation}
    a_{q_k}^\dagger(c) = |0\rangle\langle0|\otimes a_{q_k-1}^\dagger(c) + \sqrt{c+2^{q_k-1}}|1\rangle\langle0|\otimes (|0\rangle\langle1|)^{\otimes(q_k-1)} + |1\rangle\langle1|\otimes a_{q_k-1}^\dagger(c + 2^{q_k-1}).
    \label{eqn:adag_of_c_recursion}
\end{equation}

\begin{enumerate}
    \item \textbf{Base Case:} We will take $q_k=1$ for our base case. Placing this in Eq.~\ref{eqn:adag_of_c_recursion} yields

    \begin{equation}
        \begin{aligned}
            a_1^\dagger(c) &= |0\rangle\langle0|\otimes a_{0}^\dagger(c)+ \sqrt{c+2^0}|1\rangle\langle0|\otimes (|0\rangle\langle1|)^{\otimes(0)} + |1\rangle\langle1|\otimes a_{0}^\dagger(c + 2^{0})\\
            &=\sqrt{c+2^0}|1\rangle\langle0|,
        \end{aligned}
        \label{eqn: proof base case}
    \end{equation}

    which is precisely equivalent to Eq.~\ref{eqn: adag1 of c}.

    \item \textbf{Inductive Hypothesis:} Assume the recursion relation is true for some value $q_k=\xi$,

    \begin{equation}
        a_{\xi}^\dagger(c) = |0\rangle\langle0|\otimes a_{\xi-1}^\dagger(c) + \sqrt{c+2^{\xi-1}}|1\rangle\langle0|\otimes (|0\rangle\langle1|)^{\otimes(\xi-1)} + |1\rangle\langle1|\otimes a_{\xi-1}^\dagger(c + 2^{\xi-1}).
        \label{eqn: adag of c recursion}
    \end{equation}

    \item \textbf{Inductive Step:} 
     With the previous assumption, we must show that the recursion relation holds for $\xi+1$. We begin with the definition given in Eq.~\ref{eqn:adag_of_c}

    \begin{equation}
        \begin{aligned}
            a_{\xi+1}^\dagger(c) &= \sum_{n=0}^{2^{\xi+1}-2}\sqrt{c+(n+1)}|n+1\rangle_{(\xi+1) (\xi+1)}\langle n|\\
            &= \sum_{n=0}^{2^{\xi+1}-2}\sqrt{c+(n+1)}|(n+1)_{\text{msb}}\rangle\langle n_{\text{msb}}|\otimes|n_r+1\rangle_{\xi\xi}\langle n_r|.
        \end{aligned}
        \label{eqn: inductive step pt. 1}
    \end{equation}

    We have redefined the binary Fock state $|n\rangle_{(\xi+1)} = |n_{\text{msb}}\rangle\otimes|n_r\rangle_{\xi}$ where $|n_{\text{msb}}\rangle\in\{|0\rangle,|1\rangle\}$ is the most significant bit in the binary representation of $|n\rangle$, and $|n_r\rangle_\xi$ is the binary string remaining after the most significant bit is removed. In this way $n = n_r + 2^\xi n_{\text{msb}}$.

    For clarity, we show a few key values of the binary Fock state representation.

    \begin{equation}
        \begin{aligned}
            &|2^{\xi+1}-1\rangle_{\xi+1} =  |\underbrace{111\cdots11}_{\mathclap{\text{length $\xi+1$}}}\rangle =|1\rangle\otimes |\underbrace{11\cdots11}_{\mathclap{\text{length $\xi$}}}\rangle=|1\rangle\otimes\ket{2^\xi-1}_\xi,\\
            &|2^{\xi+1}-2\rangle_{\xi+1} =  |\underbrace{111\cdots10}_{\mathclap{\text{length $\xi+1$}}}\rangle=|1\rangle\otimes\ket{2^\xi-2}_\xi\\
            &\quad\quad\quad\quad\vdots\\
            &|2^{\xi}\rangle_{\xi+1} =  |\underbrace{100\cdots00}_{\mathclap{\text{length $\xi+1$}}}\rangle =|1\rangle\otimes |\underbrace{00\cdots00}_{\mathclap{\text{length $\xi$}}}\rangle=|1\rangle\otimes|0\rangle_\xi \\
            &|2^{\xi}-1\rangle_{\xi+1} =  |\underbrace{011\cdots11}_{\mathclap{\text{length $\xi+1$}}}\rangle =|0\rangle\otimes |\underbrace{11\cdots11}_{\mathclap{\text{length $\xi$}}}\rangle=\ket{0}\otimes|2^\xi-1\rangle_\xi \\
            &|2^{\xi}-2\rangle_{\xi+1} =  |\underbrace{011\cdots10}_{\mathclap{\text{length $\xi+1$}}}\rangle=|0\rangle\otimes|2^\xi - 2\rangle_\xi
        \end{aligned}
        \label{eqn: binary rep}
    \end{equation}

    It is trivially true that there are only three possible combinations of $|(n+1)_{\text{msb}}\rangle\langle n_{\text{msb}}|$, namely $|0\rangle\langle0|$, $|1\rangle\langle0|$ and $|1\rangle\langle1|$. The other permutation will not arise since the most significant bit of $|n+1\rangle$ must always be greater than or equal to the most significant bit of $|n\rangle$.

    We expand Eq.~\ref{eqn: inductive step pt. 1} into three groups corresponding to these three combinations.

    \begin{equation}
        \begin{aligned}
            a_{\xi+1}^\dagger(c) &=  |0\rangle\langle0|\otimes\sum_{n_r=0}^{2^{\xi}-2}\sqrt{c+(n+1)}|n_r+1\rangle_{\xi\xi}\langle n_r| \\
            &\quad+ \sqrt{c+2^\xi}|2^\xi\rangle_{(\xi+1)(\xi+1)}\langle2^\xi-1| \\
            &\quad +|1\rangle\langle1|\otimes\sum_{n=2^\xi}^{2^{\xi+1}-2}\sqrt{c+(n+1)}|n_r+1\rangle_{\xi\xi}\langle n_r| \\
            &=\ket{0}\bra{0}\otimes a_\xi^\dagger(c) + \sqrt{c+2^\xi}|1\rangle\langle0|\otimes(|0\rangle_{\xi\xi}\langle2^\xi-1|) \\
            &\quad +|1\rangle\langle1|\otimes\sum_{n_r=0}^{2^{\xi}-2}\sqrt{c+(n_r+2^\xi+1)}|n_r+1\rangle_{\xi\xi}\langle n_r|\\
            &=|0\rangle\langle0|\otimes a_\xi^\dagger(c) + \sqrt{c+2^\xi}|1\rangle\langle0|\otimes(|0\rangle\langle1|)^{\otimes \xi}  +|1\rangle\langle1|\otimes a_\xi^\dagger(c+2^\xi) \boxed{}
        \end{aligned}
        \label{eqn: inductive step pt. 2}
    \end{equation}

    In the last steps of Eq.~\ref{eqn: inductive step pt. 2} we have used $\ket{0}_\xi = \ket{0}^{\otimes \xi}$, $\ket{2^{\xi}-1}_\xi = \ket{1}^{\otimes \xi}$ and $n = n_r + 2^\xi n_{\text{msb}}$.

    This proves the recursion relation.
\end{enumerate}

Using the mappings

\begin{equation}
    \begin{aligned}
        |0\rangle\langle0| &= \frac{1}{2}(I+Z), &|0\rangle\langle1| = \frac{1}{2}(X+iY), \\
        |1\rangle\langle0| &= \frac{1}{2}(X-iY)\text{, and} &|1\rangle\langle1 | =\frac{1}{2}(I-Z)
    \end{aligned}
    \label{eqn:mappings}
\end{equation}

we obtain the expression for the recursion relation in terms of Pauli operators/logic gates:

\begin{equation}
    a_{q_k}^\dagger(c) = \frac{I+Z}{2}\otimes a_{q_k-1}^\dagger(c) + \frac{\sqrt{c + 2^{q_k - 1}}}{2^{q_k}}(X-iY)\otimes(X+iY)^{\otimes(q_k - 1)} + \frac{I-Z}{2}\otimes a_{q_k-1}^\dagger(c + 2^{q_k - 1}).
    \label{eqn:adag_of_c_paulis}
\end{equation}

Finally, we define the annihilation and number operators as

\begin{equation}
    \begin{aligned}
        a_{q_k}(c) &= \sum_{n=1}^{2^{q_k}-1}\sqrt{c+n}|n-1\rangle\langle n|,\quad\text{and}\\
        a^\dagger_{q_k}(c)a_{q_k}(c) &= n_{q_k}(c) = \sum_{n=1}^{2^{q_k}-1}(c+n)|n\rangle\langle n|.
    \end{aligned}
    \label{eqn: a and n defs}
\end{equation}

It is then a simple exercise to prove by induction the recursion relations

\begin{align}
    a_{q_k}(c) &= |0\rangle\langle0|\otimes a_{q_k-1}(c) + \sqrt{c+2^{q_k-1}}|0\rangle\langle1|\otimes (|1\rangle\langle0|)^{\otimes(q_k-1)} + |1\rangle\langle1|\otimes a_{q_k-1}(c + 2^{q_k-1})
    \label{eqn:a_of_c_recursion} \\
    n_{q_k}(c) &= |0\rangle\langle0|\otimes n_{q_k-1}(c) + \left(2^{q_k-1}+c\right)|1\rangle\langle1|\otimes(|0\rangle\langle0|)^{\otimes(q_k-1)} + |1\rangle\langle1|\otimes n_{q_k-1}(c+2^{q_k-1})
    \label{eqn:n_of_c_recursion}
\end{align}

or equivalently

\begin{align}
    a_{q_k}(c) &= \frac{I+Z}{2}\otimes a_{q_k-1}(c) + \frac{\sqrt{c + 2^{q_k - 1}}}{2^{q_k}}(X+iY)\otimes(X-iY)^{\otimes(q_k - 1)}  + \frac{I-Z}{2}\otimes a_{q_k-1}(c + 2^{q_k-1})
    \label{eqn:a_of_c_recursion_paulis} \\
    n_{q_k}(c) &= \frac{I+Z}{2}\otimes n_{q_k-1}(c) + \frac{2^{q_k-1}+c}{2^{q_k}}(I-Z)\otimes(I+Z)^{\otimes(q_k-1)} + \frac{I-Z}{2}\otimes n_{q_k-1}(c+2^{q_k-1}).
    \label{eqn:n_of_c_recursion_paulis}
\end{align}

Eq.~\ref{eqn:a_of_c_recursion} is simply the Hermitian conjugate of Eq.~\ref{eqn:adag_of_c_recursion}, as Eq.~\ref{eqn:a_of_c_recursion_paulis} is the Hermitian conjugate of Eq.~\ref{eqn:adag_of_c_paulis}.

\end{document}